# Multiple Source Dual Fault Tolerant BFS Trees [*]

Manoj Gupta [†]    Shahbaz Khan [‡§]

## Abstract

Let $G = (V, E)$ be a graph with $n$ vertices and $m$ edges, with a designated set of $\sigma$ sources $S \subseteq V$. The *fault tolerant subgraph* for any graph problem maintains a sparse subgraph $H = (V, E')$ of $G$ with $E' \subseteq E$, such that for any set $F$ of $k$ failures, the solution for the graph problem on $G \setminus F$ is maintained in its subgraph $H \setminus F$. We address the problem of maintaining a fault tolerant subgraph for computing *Breath First Search tree* (BFS) of the graph from a single source $s \in V$ (referred as $k$ FT-BFS) or multiple sources $S \subseteq V$ (referred as $k$ FT-MBFS). We simply refer to them as FT-BFS (or FT-MBFS) for $k = 1$, and dual FT-BFS (or dual FT-MBFS) for $k = 2$.

The problem of $k$ FT-BFS was first studied by Parter and Peleg [ESA13]. They designed an algorithm to compute FT-BFS subgraph of size $O(n^{3/2})$. Further, they showed how their algorithm can be easily extended to FT-MBFS requiring $O(\sigma^{1/2} n^{3/2})$ space. They also presented matching lower bounds for these results. The result was later extended to solve dual FT-BFS by Parter [PODC15] requiring $O(n^{5/3})$ space, again with matching lower bounds. However, their result was limited to only edge failures in undirected graphs and involved very complex analysis. Moreover, their solution doesn't seems to be directly extendible for dual FT-MBFS problem.

We present a similar algorithm to solve dual FT-BFS problem with a much simpler analysis. Moreover, our algorithm also works for vertex failures and directed graphs, and can be easily extended to handle dual FT-MBFS problem, matching the lower bound of $O(\sigma^{1/3} n^{5/3})$ space described by Parter [PODC15]. The key difference in our approach is a much simpler classification of path interactions which formed the basis of the analysis by Parter [PODC15]. Our dual FT-MBFS structure also seamlessly gives a dual fault tolerant spanner with additive stretch of $+2$ having size $O(n^{7/8})$.

## 1 Introduction

Graph networks are extensively used to study real world applications ranging from communication networks as internet and telephony, to supply chain networks, road networks etc. Every now and then, these networks are susceptible to failures of links and nodes, which drastically affects the performance of these applications. Hence, most algorithms developed for these applications are also studied in the *fault tolerant model*, which aims to provide solutions to the corresponding problem that are resilient to such failures. Since such failures of nodes or links in the network though unpredictable are rare and are often readily repaired, the applications generally address the scenarios expecting the number of simultaneous faults to be much smaller than the size of the network. This aspect is often modeled by bounding such failures using some parameter $k$ (typically $k << n$), and studying fault tolerant structures resilient to upto $k$ failures.

Among the different approaches to develop fault tolerance in a structure, we use the approach of computing a *fault tolerant subgraph* described as follows. For a given graph $G = (V, E)$, the fault tolerant subgraph for any graph problem maintains a sparse subgraph $H = (V, E')$ of $G$ having $E' \subseteq E$, such that for any set of edge (or vertex) failures $F \subseteq E$ (or $F \subseteq V$), the solution for the graph problem on $G' = (V, E \setminus F)$ (or $G' = (V \setminus F, E)$) is maintained in its subgraph $H' = (V, E \setminus F)$ (or $H' = (V \setminus F, E)$). We shall henceforth abuse the notation to denote the graphs after such a set of failures $F$ as $G \setminus F$ and $H \setminus F$ respectively. A standard motivation for this approach is a communication network where each link corresponds to a communication channel [16], where the system designer is required to purchase or lease the channels to be used by the application. Hence, the aim is to acquire a minimal set of these channels (the subgraph $H$ of $G$) for successfully performing the application with resilience of upto $k$ faults. Fault tolerant subgraphs are also developed for other graph problems maintaining reachability [13, 2, 3], strong-connectivity [3] and approximate shortest paths from a single source [12, 17, 5] and all sources [8, 10, 7, 14, 4].

---

[*] The preliminary version of this paper have been accepted to appear at ICALP 2017
[†] IIT Gandhinagar, Gandhinagar-382355, India, email: gmanoj@iitgn.ac.in
[‡] Dept. of CSE, IIT Kanpur, India, email: shahbazk@cse.iitk.ac.in
[§] This research work was supported by Google India under the Google India PhD Fellowship Award.

Breadth First Search (BFS) is a fundamental technique for graph traversal. From any given source $s \in V$, BFS produces a rooted spanning tree in $O(m + n)$ time. For an unweighted graph, the BFS tree from a source $s$ is also the shortest path tree from $s$ because it preserves the shortest path from $s$ to every vertex $v \in V$ that is reachable from $s$. We are thus interested to maintain fault tolerant subgraphs for computing BFS trees from a single source (referred as $k$ FT-BFS) and multiple sources $k$ FT-MBFS described as follows.

**Definition 1** ($k$ FT-BFS)**.** *Given a graph $G = (V, E)$ with a designated source $s \in V$, build a subgraph $H = (V, E')$ with $E' \subseteq E$, such that after any set $F$ of $k$ failures in $G$, the BFS tree from $s$ in $H \setminus F$ is a valid BFS tree from $s$ in $G \setminus F$.*

**Definition 2** ($k$ FT-MBFS)**.** *Given a graph $G = (V, E)$ with a designated set of sources $S \subseteq V$, build a subgraph $H = (V, E')$ with $E' \subseteq E$, such that after any set $F$ of $k$ failures in $G$, for each $s \in S$ the BFS tree from $s$ in $H \setminus F$ is a valid BFS tree from $s$ in $G \setminus F$.*

For convenience of notation, for $k = 1$ and $k = 2$ we refer to these problems as FT-BFS (or FT-MBFS) and dual FT-BFS (or dual FT-MBFS). The problems of $k$ FT-BFS (and $k$ FT-MBFS) were first studied by Parter and Peleg [16] for a single failure. They designed an algorithm to compute FT-BFS requiring $O(n^{3/2})$ space. Further, they showed their result can be easily extended to FT-MBFS requiring $O(\sigma^{1/2} n^{3/2})$ space. Moreover, their upper bounds were complemented by matching lower bounds for both their results. This result was later extended to address dual FT-BFS by Parter [15] requiring $O(n^{5/3})$ space. However, the application of this result was limited to only edge failures in undirected graphs. Though the analysis of their result was significantly complex, it paved a way for developing the theory studying the interaction of replacement paths after a single edge failure, their classification and corresponding properties. Further, they also generalized the lower bound for $k$ FT-MBFS to $\Omega(\sigma^{\frac{1}{k+1}} n^{2-\frac{1}{k+1}})$ which matches their solution for dual FT-BFS. They also stated extensions of their result to dual FT-MBFS (or $k$ FT-BFS) as an open problem.

The difference in complexity of dual FT-BFS over FT-BFS also reinforces the idea that extending such results from one failure to two failures (and beyond) requires a significantly more advanced analysis. As described by Parter [15], for the problem of maintaining shortest paths "*a sharp qualitative and quantitative difference*" has been widely noted while handling a single failure and multiple failures. For the problem of maintaining fault tolerant distance oracles, despite a simple and elegant algorithm for a single edge failure [9], the solution for two edge failures [11] is significantly complex. In fact, the authors [11] themselves mention that extending their approach beyond 2 edge failure would be infeasible due to numerous case analysis involved, requiring a fundamentally different approach. This key difference is also visible when we compare other problems, as bi-connectivity with tri-connectivity, single fault tolerant reachability [13, 2] with dual fault tolerant reachability [3], etc. Hence, simplifying the analysis of dual FT-BFS (and hence dual FT-MBFS) structures seem to be an essential building block for further developments of the problem for multiple failures.

## 1.1 Our Contributions

We design optimal algorithms for constructing dual FT-BFS and dual FT-MBFS structures. In principle, the core algorithm of our construction for dual FT-BFS is same as the one given by Parter [15], with a much simpler and more powerful analysis. As a result, our algorithm also works for vertex failures and directed graphs. Also, our dual FT-BFS structure can also be easily extended to handle dual FT-MBFS (as in case of FT-BFS [16]), which matches the lower bound described by Parter [15]. Thus, we optimally solve two open problems (dual FT-BFS for directed graphs and dual FT-MBFS for any graphs) as follows.

**Theorem 3** (Optimal dual FT-BFS). *Given any graph $G = (V, E)$ having $n$ vertices and $m$ edges, with a designated source $s \in V$, there is a polynomial time constructable dual FT-BFS subgraph $H$ having $O(n^{5/3})$ edges.*

**Theorem 4** (Optimal dual FT-MBFS). *Given any graph $G = (V, E)$ having $n$ vertices and $m$ edges, with a designated set of $\sigma$ sources $S \subseteq V$, there is a polynomial time constructable dual FT-MBFS subgraph $H$ having $O(\sigma^{1/3} n^{5/3})$ edges.*

Our analysis is performed using simple techniques based on *counting arguments*. We classify a set of shortest paths as *standard* paths and prove the properties of *disjointness* and *convergence* for a designated suffix of such paths. The extension to directed graphs additionally uses the notion of *segmentable* paths (similar notion of *regions* was used in [15]) for every set of *converging* shortest paths, and establishes several interesting properties for them. These properties and analysis techniques might be of independent interest in the theory of shortest paths. Finally, using standard constructions [14, 4] our dual FT-MBFS structure can be seamlessly used to build a dual fault tolerant spanner with additive stretch 2 requiring $O(n^{7/8})$ edges.

## 1.2 Related Work

As described earlier BFS is strongly related to shortest paths. Demetrescu et.al. [9] showed that there exist weighted directed graphs, for which a fault tolerant subgraph requires $\Theta(m)$ edges for maintaining shortest paths even from a single source after a vertex failure. Hence, they designed a data-structure of size $\tilde{O}(n^2)$ [1] that reports all pairs shortest distances after a vertex failure in $O(1)$ time. Duan and Pettie [11] extended this result to two failures requiring nearly same (upto $poly \log n$ factors) size and reporting time.

Other related problems include fault tolerant DFS and fault tolerant reachability. Baswana et al. [1] presented a $\tilde{O}(m)$ sized fault tolerant data structure that reports the DFS tree of an undirected graph after $k$ faults in $\tilde{O}(nk)$ time. For single source reachability, Baswana et al. [3] presented an algorithm for computing fault tolerant reachability subgraphs for $k$ faults using $O(2^k n)$ edges. This result was also shown to be optimal upto constant factors.

### Outline of the paper

We now present a brief outline of our paper. In Section 2, we present the basic notations that shall be used throughout the paper, which shall be followed by a brief overview of our approach and analysis in Section 3. For the sake of simplicity we first describe our analysis for undirected graphs. In Section 4, we shall first begin with the description of our algorithm for dual FT-BFS and the properties of the shortest paths found using it, which shall be followed by the formal analysis. We then present our algorithm for dual FT-MBFS and its analysis, drawing similarities with solution of dual FT-BFS. In Section 6 we extend this analysis for directed graphs. Section 7 describes how our dual FT-MBFS structure can be used to build a dual fault tolerant spanner with additive stretch 2. Finally, we present the concluding remarks for our paper in Section 8. In the interest of completeness, some previously proved results used by our paper have been proved using simpler techniques in Appendix. For the sake of simplicity, we only describe our algorithm and analysis for edge failures. However, the same analysis can also be used to handle vertex failures.

---

[1] $\tilde{O}(\cdot)$ notation hides poly-log($n$) factors

## 2  Preliminaries

Given a graph $G = (V, E)$ with $n$ vertices and $m$ edges with a set of designated source $s \in S$. The following notations shall be used throughout the paper.

- $P, \mathcal{P}$: A path is denoted by $P$, where $Source(P)$ and $Dest(P)$ represents the source and destination of path $P$. In most parts of the paper, $Source(P) = s$ and $Dest(P) = v$. A set of paths is denoted by $\mathcal{P}$. Generally, we assume a path from $s$ to $v$ starts from the top ($s$) and ends at bottom ($v$). For two paths $P', P''$, we say $P'$ leaves *earlier/higher* (or *later/lower*) than $P''$ from $P$, if $P'$ leaves $P$ closer to $s$ (or closer to $v$) than $P''$.

- $F(P)$: For the shortest path $P$ from $Source(P)$ to $Dest(P)$ after a set of edge failures, this set of failed edges is denoted by $F(P) = \{e_1, e_2, \ldots, e_k\}$ (say), where $e_i$ denotes the $i^{th}$ edge in the sequence. Similarly for some path $P'$, $e'_i$ denotes the $i^{th}$ edge in the sequence.

- $P_i$: If $F(P) = \{e_1, e_2, \ldots, e_k\}$, then $P_i$ is the shortest path avoiding the first $i$ edge of $F(P)$, i.e., $F(P_i) = \{e_1, e_2, \ldots, e_i\}$, where $0 \le i < k$. Again, for most parts of the paper, $P_0$ denotes the shortest path from $s$ to $v$ in $G$.

- $D_i(P)$: If $|F(P)| = k$, the detour path of $P$ from $P_i$, $D_i(P) = P \setminus \{\cup_{j=0}^{i} P_j\}$ [2], where $1 \le i < k-1$. For dual case, $D_0(P)$ is the detour of $P$ from $P_0$, $D_1(P)$ is the detour of $P$ from $P_1$, and $D_0(P_1)$ is the detour of $P_1$ from $P_0$ ( See Figure 1).

- $LastE(P)$ : The last edge of a path $P$.

- $P[x, y]$: The sub-path of $P$ starting from $x$ to $y$, where $x, y \in P$.

We define the property of *convergence* of a set of paths $\mathcal{P}$ as follows. The paths in $\mathcal{P}$ are said to be *converging* if on intersection of any two paths $P, P' \in \mathcal{P}$, both $P$ and $P'$ merge and do not diverge till the end of the paths.

## 3  Overview

For analyzing the size of dual FT-BFS subgraph, i.e., the number of edges in shortest paths from the source $s$ to each vertex $v \in V$ after any two failures, it suffices to count only the last edge of every such path $P$, for each $v \in V$ [16, 15]. The novelty of our approach is the classification of such paths based on interaction of corresponding $P_1$ and $P_0$, whereas Parter [15] studied the different interactions of $P_1$ and $P'_1$, for two such paths $P$ and $P'$.

We primarily use the disjointness of a designated suffix of such a path $P$ (referred as $LastLeg(P)$) with *counting* arguments to bound the number of such paths. To achieve this, we classify some of these paths as *standard paths* based on the interactions of corresponding $P_1$ and $P_0$. The number of *non-standard* paths can be easily bound using simple *counting* arguments. The set of *standard* paths exhibit several interesting properties including convergence of corresponding paths $D_0(P_1)$. We further classify the *standard* paths into *long standard* paths and *short standard* paths, each bounded separately using relatively harder techniques. For sake of easier presentation we first bound the number of *short standard* paths only for undirected graphs, with extension to directed graphs requiring an additional notion of *segmentable* paths. The only difference in the analysis of dual FT-MBFS is the definition of *standard* paths and dealing with interaction of $P_1$ with $P_0'$ corresponding to other sources.

---
[2] This construction may give a set of disjoint subpaths of $P$ instead of a single subpath. However, in most cases this path will be a single subpath, else we assume $D_i(P)$ to be the last such subpath on $P$.

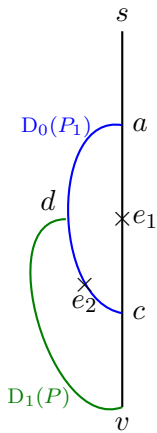

Figure 1: Showing $P_0$ (in black), $D_0(P_1)$ (in blue) and $D_1(P)$ (in green). Here $P_1 = P_0[s,a] \cup D_0(P_1) \cup P_0[c,v]$ and $P = P_0[s,a] \cup D_0(P_1)[a,d] \cup D_1(P)$.

## 4 Dual FT-BFS

We shall now describe our algorithm to compute sparse dual FT-BFS subgraph $H$ from a source $s \in V$. For every vertex $v \in V$, our algorithm computes the shortest paths from $s$ to $v$ avoiding upto two failures and adds the last edge of each such path to the adjacency list of vertex $v$. Note that repeating the procedure for each vertex on such a path adds the entire path to $H$ [16, 15].

Our algorithm starts by adding the shortest path between $s$ and $v$, i.e., $P_0$. It then processes single edge failures on $P_0$. We then find the replacement path $P$ for all two edge failures $\{e_1, e_2\}$ such that $e_1 \in P_0$ and $e_2 \in P_1$. Further, in case $e_2 \in P_0 \cap P_1$ then $e_1$ is higher than $e_2$ on $P_0$.

However, we want to process all the failures in some particular order. This ordering plays a crucial role in the analysis. To this end, we define this ordering $\pi$ as follows. The first failure in $\pi$ is $F = \emptyset$, which adds $P_0$. The ordering $\pi$ then contains single edge failures of type $F = \{e\}$ (where $e \in P_0$), ordered by their decreasing distance from $s$ on $P_0$. Finally, we order the remaining failures as follows: for any two failures $F = \{e_1, e_2\}$ and $F' = \{e_1', e_2'\}$ (with corresponding replacement paths $P$ and $P'$), $F \prec_\pi F'$ if either (1) $e_1$ is farther than $e_1'$ from $s$ on $P_0$, or (2) $e_1 = e_1'$ and $e_2$ is farther than $e_2'$ from $s$ on $P_1$ (note that $P_1 = P_1'$ in this case). If $F \prec_\pi F'$, $F$ is said to be *lower* than $F'$ in $\pi$.

For any failure of $F = \{e_1, \cdots, e_k\}$, we define the *preferred* shortest path avoiding $F$. Our preferred shortest path will be a path of shortest length avoiding $F$. However, there can be multiple such paths of same length. We use following rules to choose a unique preferred path.

**Definition 5.** *Path $P$ is **preferred** for failure of $\{e_1, \cdots, e_k\}$ where each $e_i \in P_{i-1}$, if*

1. *For each $i$, $P$ leaves $P_{i-1}$ before $e_i$ exactly once.*

2. For any other $P'$ also avoiding $\{e_1, \cdots, e_i\}$, we have either (i) $|P| < |P'|$, (ii) $|P| = |P'|$, and for some $0 \le i \le k$, both $P$ and $P'$ leaves each of $P_0, ..., P_{i-1}$ at the same vertex, but $P$ leaves $P_i$ earlier than $P'$, (iii) $P$ is lexicographically smaller [3] than $P'$.

Intuitively, out of all the shortest paths avoiding $F$ (say for $|F| = 2$), the preferred path leaves the path $P_0$ and/or $P_1$ as early as possible. In order to avoid the preferred path leaving $P_0$ (or $P_1$) multiple times just to achieve an earlier point of divergence from $P_0$ (or $P_1$), the first condition is imposed. The last condition in (2) is just to break ties between two paths that are of same length and leave $P_0$ and $P_1$ at the same vertex.

Finally, in order to add the preferred shortest path $P$ avoiding a failure $F$, our algorithm simply adds $LastE(P)$ to $H$, which suffices to add the entire path as described earlier. Moreover, we also assign the corresponding $P$ to the failure $F$ if it was the first failure to add this edge in $H$. As a result, if $P$ and $P'$ are two preferred paths avoiding $F$ and $F'$ respectively where $LastE(P) = LastE(P')$, then if $F \prec_\pi F'$, only the path $P$ shall be assigned to $F$. Refer to Procedure Dual-FT-BFS for the pseudocode of our algorithm.

---

**Procedure** Dual-FT-BFS$(s, v, \pi)$: Augments the dual FT-BFS subgraph $H$, such that for BFS tree of $G$ rooted at $s$ after any two edge failures in $G$, the incoming edges to $v$ are preserved in $H$.

1 **foreach** *Failure $F$, where $0 \le |F| \le 2$, ordered from lower to higher in $\pi$* **do**
2     $P \leftarrow$ *Preferred* path from $s$ to $v$ in $G$ avoiding $F$;
3     **if** $LastE(P) \notin H$ **then**
4        Assign $P$ for failure of $F$;
5        Add $LastE(P)$ to $H$;
6     **end**
7 **end**

---

In order to calculate the size of $H$, it is sufficient to analyze the number of different last edges added on each $v \in V$ in $H$. Let the set of all paths from $s$ to $v$ avoiding failures $F \subseteq E$ (where $|F| \le 2$) be $\mathcal{P}_v$. We thus define the paths that will be counted for establishing the space bound as follows.

**Definition 6.** *The path $P \in \mathcal{P}_v$ is called* contributing *if while processing* $F(P)$*, $LastE(P) \notin H$, i.e., $P$ adds a new edge adjacent to $v$ in $H$.*

In order to count the number of contributing paths to a vertex $v$, we only need to consider its interactions with other contributing paths in $\mathcal{P}_v$. This is because, if any other path $P \in \mathcal{P}_x$ passes through $v$ using some new edge, so does the corresponding $P' \in \mathcal{P}_v$ with $F(P) = F(P')$. Thus, to analyze the size of $H$, it suffices to look at last edges of the contributing paths in $\mathcal{P}_v$ for each vertex $v$ separately.

### 4.1 Properties of contributing paths

Parter [15] presented a simple proof bounding the number of contributing paths avoiding multiple failures on $P_0$ to $O(\sqrt{n})$ for each vertex $v$. Hence, excluding these paths, every contributing path satisfies the following properties.

**Lemma 7.** *Excluding $O(\sqrt{n})$ paths, each contributing path $P$ from $s$ to $v$ avoiding $\{e_1, e_2\}$ satisfies following properties*

---

[3] Let $P$ and $P'$ first diverge from each other to $x \in P$ and $x' \in P'$ respectively, i.e., $P[s,x] \setminus \{x\} = P'[s,x'] \setminus \{x'\}$. If the index of $x$ is lower than that of $x'$ then $P$ is said to be *lexicographically smaller* than $P'$.

$\mathbb{P}_1$ : $e_1 \in P_0$ and $e_2 \in D_0(P_1)$.

$\mathbb{P}_2$ : *Except at $v$, $D_0(P)$ does not intersect with $P_0$ and $D_1(P)$ does not intersect with $P_1$, after diverging from $P_0$ and $P_1$ respectively.*

$\mathbb{P}_3$ : *For any path $P'$ which avoids $\{e_1, e_2\}$, $P$ is the preferred path over $P'$.*

$\mathbb{P}_4$ : *If $P$ also avoids some failure $F'$ where $F' \prec_\pi F$, then there exist another path $P'$ which is the preferred path for $F'$ over $P$, where $P'$ does not avoid $F$.*

*Proof.* $\mathbb{P}_1$ : Parter [15] presented a simple proof bounding the number of contributing paths avoiding multiple failures on $P_0$ to $O(n^{3/2})$ (an alternate proof using *counting arguments* is presented in Appendix A for the sake of completeness). Hence, excluding these $O(n^{3/2})$ paths, every contributing path satisfies $\mathbb{P}_1$ since $e_2 \in P_1$.

$\mathbb{P}_2$ : In order to prove $\mathbb{P}_2$, consider a path $P$ with $D_0(P)$ intersecting $P_0$ at some vertex $w$ after diverging from $P_0$ at $x$. Since $P$ is a preferred path, it cannot leave $P_0$ more than once from $P_0$ before the failing edge $e_1$ ensuring $e_1 \notin P_0[w, v]$. Further, we also have $e_2 \notin P_0$ (by $\mathbb{P}_1$) which ensures that $P$ would continue to follow $P_0$ after $w$ ($P_0$ being the lexicographically shortest path), making $P$ non-contributing. Similarly consider a path $P$ with $D_1(P)$ intersecting $P_1$ at some vertex $w$ after diverging from $P_1$ at $x$. Again, our algorithm ensures that $P$ cannot leave more than once from $P_1$ before the failing edge $e_2$ ensuring $e_2 \notin P_1[w, v]$. Thus, $P$ would continue to follow $P_1$ after $w$ making it non-contributing.

$\mathbb{P}_3$ : This property directly follows from Procedure Dual-FT-BFS by construction.

$\mathbb{P}_4$ : In order to prove $\mathbb{P}_4$, clearly since $F'$ was processed before $F$ by our algorithm it cannot have the preferred path $P$ else $P$ will be associated with failure $\{e'_1, e'_2\}$ and not $\{e_1, e_2\}$. Moreover, the preferred path $P'$ for $\{e'_1, e'_2\}$ cannot avoid $\{e_1, e_2\}$ else it would also be the preferred path for $\{e_1, e_2\}$ (and not $P$). □

## 4.2 Space Analysis

As described earlier, in order to bound the size of dual FT-BFS subgraph to $O(n^{5/3})$, it suffices to bound the number of *contributing* paths from $s$ to each vertex $v \in V$ avoiding two edge failures to $O(n^{2/3})$. Further, using $\mathbb{P}_1$ we are only concerned with a contributing path $P$ if $e_1 \in P_0$ and $e_2 \in D_0(P_1)$.

We first divide the path $P_0$ into two parts as follows. Let $v_l \in P_0$ be the vertex such that $|P_0[v_l, v]| = n^{1/3}$. We define $P_{high} = P_0[s, v_l]$ and $P_{low} = P_0[v_l, v]$. If $|P_0| < n^{1/3}$, we assume $v_l = s$ where $P_{high} = \phi$. We shall now define the *standard paths* as follows.

**Definition 8** (Standard Paths). *A contributing path $P$ is called a* standard path *if (a) $e_1 \in P_{high}$, and (b) $D_0(P_1)$ merges with $P_0$ on $P_{low}$, i.e., $Dest(D_0(P_1)) \in P_{low}$.*

We can thus classify the contributing paths into following three types (see Figure 2):

$\mathscr{P}_a$: Non-standard paths.

$\mathscr{P}_b$: Long standard paths, i.e., standard paths with $|D_0(P_1)| \geq n^{2/3}$.

$\mathscr{P}_c$: Short standard paths, i.e., standard paths with $|D_0(P_1)| < n^{2/3}$.

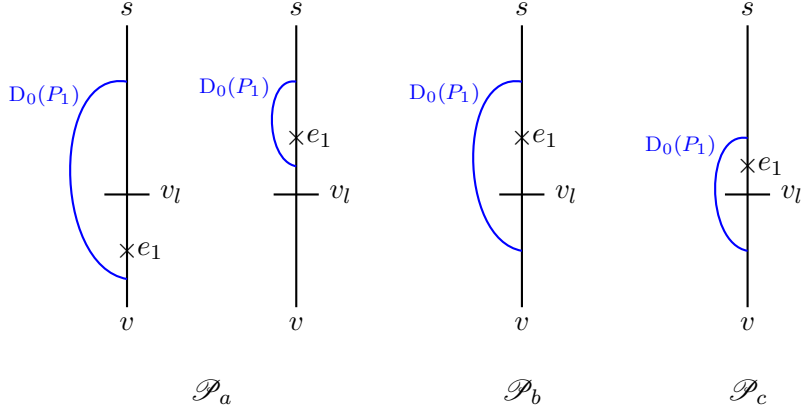

Figure 2: Classification of contributing paths:
$\mathscr{P}_a$: Non-Standard Paths, $\mathscr{P}_b$ : Long Standard Paths and $\mathscr{P}_c$ : Short Standard Paths.

Clearly, the sets $\mathscr{P}_a, \mathscr{P}_b$ and $\mathscr{P}_c$ are mutually disjoint and collectively exhaustive. Further, we define a set $\mathscr{P}_{1x}$ (for $x = a, b$ and $c$), where for each $P \in \mathscr{P}_x$, we add the corresponding $P_1$ to $\mathscr{P}_{1x}$. In addition, we identify the disjoint suffix of a path $P$ as follows (see Figure 3).

**Definition 9.** *For each $P \in \mathscr{P}_x$, for $x = a, b$ or $c$, we define the following*

1. *LastPath(P) : The path in $\mathscr{P}_{1x}$ that intersects last with $P$. If $P$ diverges from $P_0$ and does not intersect any path in $\mathscr{P}_{1x}$, we set $LastPath(P) = P_0$.*

2. *LastLeg(P) : The part of $P$ after diverging from LastPath(P), i.e., $P[v^*, v]$, where $v^*$ is the last vertex of $P$ on $P \cap LastPath(P)$.*

The suffix $LastLeg(P)$ of a contributing path $P$ satisfies the following properties.

**Lemma 10.** *For every set $\mathscr{P}_x$ (for $x = a, b$ or $c$), we have the following.*

a. *For any $P, P' \in \mathscr{P}_x$, LastLeg(P) and LastLeg(P') are disjoint (except at $v$), i.e., $LastLeg(P) \cap LastLeg(P') = \{v\}$. Further, each $P, P'$ starts from a distinct vertex on $\mathscr{P}_{1x}$.*

b. *Number of paths $P \in \mathscr{P}_x$ with $|LastLeg(P)| > n^{1/3}$ or $LastPath(P) = P_0$, is $O(n^{2/3})$.*

*Proof.* a. Consider two paths $P, P' \in \mathscr{P}_x$. For contradiction, assume that $w$ is the last vertex at which $LastLeg(P)$ intersects $LastLeg(P')$. This implies that $P[w, v]$ and $P'[w, v]$ are vertex disjoint except at $v$ and $w$. Since both $P$ and $P'$ are preferred paths, it is only possible if $P[w, v]$ passes through either $e'_1$ or $e'_2$ (or $P'[w, v]$ passes through $e_1$ or $e_2$). Assume that $P[w, v]$ passes through either $e'_1$ or $e'_2$ (the second case is identical to this case). In that case, a portion of last leg of $P$, that is $P[w, v]$ intersects either $P_0$ (since $e'_1 \in P_0$) or $\mathscr{P}_{1x}$ (since $e'_2 \in P'_1$ and $P'_1 \in \mathscr{P}_{1x}$). This contradicts the definition of $LastLeg(P)$. Hence, $LastLeg(P)$ and $LastLeg(P')$ cannot diverge after intersecting.

Now, we will use the this property to show that $LastLeg(P)$ and $LastLeg(P')$ are disjoint. Each contributing path $P$ contributes a different last edge $LastE(P)$ incident on $v$. Thus, $LastLeg(P)$ and $LastLeg(P')$ do not intersect except at $v$. This also ensures that $Source(LastLeg(P)) \neq Source(LastLeg(P'))$ proving the claim.

b. Consider any path $P \in \mathscr{P}_x$ with $|LastLeg(P)| > n^{1/3}$. Using (a), for each such $P$ we can associate $|LastLeg(P)| > n^{1/3}$ unique vertices of $LastLeg(P)$ from a total of at most $n$ vertices. Thus, the number of such paths is $O(n^{2/3})$. Similarly, consider any path $P \in \mathscr{P}_x$ where $LastLeg(P)$ diverges from $P_0$, i.e., $LastPath(P) = P_0$. Using (a), each such path emerges from a different vertex on $P_0$, limiting the number of paths emerging from $P_{low}$ to $O(n^{1/3})$. For the remaining paths, the corresponding $LastLeg(P)$ are at least as long as $P_{low}$ or $n^{1/3}$, limiting them to $O(n^{2/3})$ as described above.

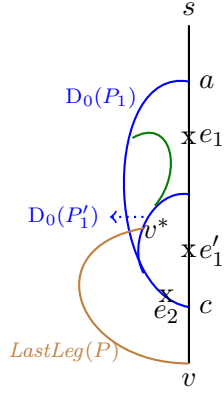

Figure 3: $P$ avoids $\{e_1, e_2\}$. Its detour $D_1(P)$ (shown in blue) last intersects $LastPath(P) = P_1'$. $P$ diverges from $P_1'$ at $v^*$, i.e., $LastLeg(P) = P[v^*, v]$ (shown in brown).

□

**Remark:** Lemma 10b claims that $LastLeg(P)$ is disjoint from other $LastLeg(P')$, where $P \in \mathscr{P}_x$ and $P' \in \mathscr{P}_{x'}$ only when $x = x'$. However, in case $x \neq x'$ they can intersect and our proof does not require their disjointness.

Equipped with these properties we can easily analyze the number of *non-standard paths* ($\mathscr{P}_a$) and *standard paths* ($\mathscr{P}_b$ and $\mathscr{P}_c$) in the following sections.

### 4.2.1 Analyzing non-standard paths $\mathscr{P}_a$

Using Lemma 10b, we know that the number of $P \in \mathscr{P}_a$ with $|LastLeg(P)| > n^{1/3}$ or $LastPath(P) = P_0$ is $O(n^{2/3})$. We now focus on the case when $|LastLeg(P)| \leq n^{1/3}$ and $LastPath(P) \in \mathscr{P}_{1a}$. For any path $P$, let $v^* = Source(LastLeg(P))$. Since $LastLeg(P)$ is a detour from $LastPath(P)[v^*, v]$ avoiding the entire $P_0$ (using $\mathbb{P}_2$), we have $|LastPath(P)[v^*, v]| \leq |LastLeg(P)| \leq n^{1/3}$. By definition, a contributing path $P$ is *non-standard* if either (a) $e_1 \in P_{low}$, or (b) $D_0(P_1)$ merges with $P_0$ on $P_{high}$, i.e., $Dest(D_0(P_1)) \in P_{high}$. Hence, for every $P$, $LastPath(P)$ would correspond to one of the two cases (a) or (b). Case (b) is clearly not be applicable here because $|LastPath(P)[v^*, v]| \geq |P_{low}| = n^{1/3}$ (since $Dest(LastPath(P)) \in P_{high}$). For case (a), on each $LastPath(P) \in \mathscr{P}_{1a}$, $v^*$ can be one of $n^{1/3}$ vertices of $LastPath(P)$ closest to $v$. Further, since $e_1 \in P_{low}$, there are only $n^{1/3}$ such paths in $\mathscr{P}_{1a}$ because each such path corresponds to failure of unique edge in $P_{low}$. Thus, there are only $n^{1/3} \times n^{1/3} = n^{2/3}$ different vertices $v^*$ limiting the number of $P \in \mathscr{P}_a$ with $|LastLeg(P)| \leq n^{1/3}$ to $O(n^{2/3})$ (using Lemma 10a).

### Properties of standard paths ($\mathscr{P}_b$ or $\mathscr{P}_c$)

We shall now prove two important properties of standard paths. The first result states that if $D_0(P_1)$ and $D_0(P_1')$ intersect, where $P, P' \in \mathscr{P}_{1b} \cup \mathscr{P}_{1c}$, then they cannot diverge. The second result states that the length of paths in $\mathscr{P}_b \cup \mathscr{P}_c$ are different. A similar result was proved by Parter[15].

**Lemma 11.** *For the set of contributing* standard *paths, we have the following properties.*

a. *The set of paths $\{D_0(P_1)|P_1 \in \mathscr{P}_{1b} \cup \mathscr{P}_{1c}\}$, is converging.*

b. *(Parter [15]) For any two paths $P, P' \in \mathscr{P}_b \cup \mathscr{P}_c$, $|P| \neq |P'|$.*

*Proof.* a. Let $w$ be the last vertex at which $D_0(P_1)$ intersects $D_0(P_1')$. This implies that $D_0(P_1)[w, v]$ and $D_0(P_1')[w, v]$ are vertex disjoint except at $w$ and $v$. Since both $P_1$ and $P_1'$ are preferred shortest paths, this is possible only if $e_1 \in D_0(P_1')[w, v]$ or $e_1' \in D_0(P_1)[w, v]$. We deal with the first case (the second case is identical). Since both $e_1, e_1' \in P_0$, this implies that $e_1 \in D_0(P_1')[w, v] \cap P_0$. This path is a part of $P_{low}$ because $D_0(P_1')$ intersects at $P_{low}$ (recall definition of $\mathscr{P}_c$ and $\mathscr{P}_c$). This is not possible because by definition, for any path $P$ in $\mathscr{P}_c$ and $\mathscr{P}_c$, we have $e_1$ lies in $P_{high}$. Hence, by contradiction no such $w$ is possible, proving the lemma.

b. This was originally proved by Parter [15] (an alternate proof is presented in Appendix B). □

### 4.2.2 Analyzing long standard paths $\mathscr{P}_b$

We first prove a generic technique to bound the number of contributing paths $P$ if the set of corresponding paths $P_1$ is converging and each $P_1$ sufficiently long.

**Theorem 12.** *Given a set $\mathcal{P}$ of converging paths satisfying Lemma 10a, where for each $P_1 \in \mathcal{P}$ we have $|P_1| \geq \alpha^2$ (where $\alpha \geq 1$), the number of contributing paths $P$ having $P_1 \in \mathcal{P}$ is $O(n/\alpha)$.*

*Proof.* Recall the definition of $LastPath(P)$, here we shall define $LastPath(P)$ (and hence $LastLeg(P)$) corresponding to paths in $\mathcal{P}$ (rather than $\mathscr{P}_{1x}$ in Definition 9). Using Lemma 10a, if $|LastLeg(P)| \geq \alpha$, then $P$ can be associated with $\alpha$ unique vertices of $LastLeg(P)$. This limits the total number of such paths to $O(n/\alpha)$. Hence, we assume that $LastLeg(P)) \leq \alpha$.

For each path $P_1 \in \mathcal{P}$, let $v_l = Dest(P_1)$. Similarly, for each such $P$, let the last intersection vertex of $LastLeg(P)$ and $LastPath(P)$ be $v^*$. Using Lemma 10a, we know that for each such contributing path $P$, its corresponding $LastLeg(P)$ starts from a distinct vertex of $\mathcal{P}$. Since $LastLeg(P)$ is a detour from $LastPath(P)[v^*, v_l]$ avoiding the entire $P_1$ (using $\mathbb{P}_2$), we have $|LastLeg(P)| \geq |LastPath(P)[v^*, v_l]|$. Since $|LastLeg(P)| \leq \alpha$, $v^*$ can be one of $\alpha$ vertices of $LastPath(P)$ closest to $v_l$.

We shall associate each such vertex $v^*$ on $LastPath(P) \in \mathcal{P}$ uniquely with $\alpha$ vertices of $LastPath(P)$, for all $LastPath(P) \in \mathcal{P}$, as follows. Let the vertices of some $LastPath(P)$ be $v_1, ..., v_k$ where $v_1$ is the closest vertex to $v_l$. For each $v_i$, $i = 1, ..., \alpha$, we associate the vertices $v_{(i-1)\alpha}, ..., v_{i\alpha}$. Since $|LastPath(P)| \geq \alpha^2$ (by definition of $\mathcal{P}$) and $i \in [1, \alpha]$ such an association can be made. Now, in order to prove that such an association is unique, i.e., a vertex $x$ is not associated with two different vertices $v_1^*, v_2^*$ of $\mathcal{P}$, we exploit the convergence of $\mathcal{P}$ as follows. Clearly if $x \in P_1$ for a unique path $P_1 \in \mathcal{P}$, there is a unique $v_1^* \in \mathcal{P}$ to which it is associated. However, if $x \in P_1$ and $x \in P_1'$ for any two paths $P_1, P_1' \in \mathcal{P}$, then $P_1$ and $P_1'$ will not diverge after intersection (by convergence of $\mathcal{P}$). This implies $P_1[x, v_l] = P_1'[x, v_l']$. Thus, the corresponding $v_1^* \in P_1$ and $v_2^* \in P_1'$ would also be same as by definition $v_1^* \in P_1[x, v_l]$. Hence, for every $P$ emerging from $v^*$ with $|LastPath(P)[v^*, v_l]| \leq \alpha$, the corresponding $v^*$ can be uniquely associated with at least $\alpha$ vertices of $\mathcal{P}$. This limits the total number of such paths to $O(n/\alpha)$ proving the theorem. □

Using Lemma 11a and by definition of long standard paths $\mathscr{P}_b$, Theorem 12 is applicable for the set $D_0(P_1)$ for $P_1 \in \mathscr{P}_{1b}$ and $\alpha = n^{1/3}$ limiting the number of paths in $\mathscr{P}_b$ to $O(n^{2/3})$.

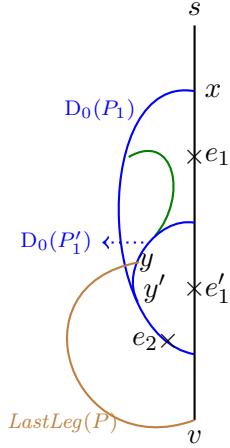

Figure 4: Let $P_1'$ be $LastPath(P)$. Then the path $P_0[s,x] \cup P_1[x,y'] \cup P_1'[y',y] \cup LastLeg(P)$ is a valid path avoiding $\{e_1, e_2\}$.

### 4.2.3 Analyzing short standard paths $\mathscr{P}_c$

To highlight the simplicity of our approach, we only analyze the paths in $\mathscr{P}_c$ for undirected graphs here. The extension of this proof for directed graphs is presented in Section 6.

Using Lemma 10b, we know that the number of $P \in \mathscr{P}_c$ with $|LastLeg(P)| > n^{1/3}$ or $LastPath(P) = P_0$ is $O(n^{2/3})$. We now focus on the case when $|LastLeg(P)| \le n^{1/3}$ and $LastPath(P) \in \mathscr{P}_{1c}$. Any such contributing path $P$ can be divided into two parts (see Figure 4), (a) $P[s,y]$, where $y = Source(LastLeg(P))$, and (b) $P[y,v] = LastLeg(P)$. We will now find an alternate path for $P[s,y]$, which will help us in bounding its length. Since $P$ is a contributing path, it diverges from $LastPath(P)$ which requires either $e_1$ or $e_2$ to be on $LastPath(P)[y,v]$. By definition of *standard paths*, we have $\mathrm{D}_0(LastPath(P))$ terminates on $P_0$ only on $P_{low}$, whereas $e_1 \notin P_{low}$ ensuring that $e_1 \notin LastPath(P)$. Thus, $e_2 \in LastPath(P)[y,v]$ and hence it intersects with $P_1$ as $e_2 \in P_1$. Using Lemma 11a, we can thus say that $LastPath(P)$ and $P_1$ merge at some vertex say $y'$, where $e_2 \in LastPath(P)[y',v] = P_1[y',v]$ (see Figure 4). We have an alternate path for $P[s,y]$ avoiding $F(P)$ formed by $P_1[s,y'] \cup LastPath(P)[y',y]$. Let $x = Source(\mathrm{D}_0(P_1))$. Since $P[s,v]$ is the shortest path avoiding $F(P)$ we have

$$\begin{aligned}
|P| &= |P[s,y]| + |P[y,v]| \\
&= |P_1[s,y]| + |P[y,v]| \\
&\le |P_1[s,y']| + |LastPath(P)[y',y]|) + |LastLeg(P)[y,v]| \\
&= (|P_1[s,x]| + |P_1[x,y']|) + |LastPath(P)[y',y]| + |LastLeg(P)[y,v]| \\
&\le |P_0| + |\mathrm{D}_0(P_1)| + |\mathrm{D}_0(LastPath(P))| + |LastLeg(P)| \\
&\le |P_0| + n^{2/3} + n^{2/3} + n^{1/3} \qquad \text{(by definition of } \mathscr{P}_c)
\end{aligned}$$

Now, using Lemma 11b, we know that for any $P, P' \in \mathscr{P}_c$ we have $|P| \ne |P'|$. We thus arrange the paths in $\mathscr{P}_c$ (except the ones in Lemma 10b) in the increasing order of sizes, where $i^{th}$ such path has the length $\ge |P_0| + i$ (as all paths at least as long as $P_0$). Since for any such $P \in \mathscr{P}_c$ we also have $|P| \le |P_0| + 3n^{2/3}$ (described above), clearly the number of paths in $\mathscr{P}_c$ are $O(n^{2/3})$ (for $i$ upto $3n^{2/3}$).

This completes the proof of our dual FT-BFS result in Theorem 3.

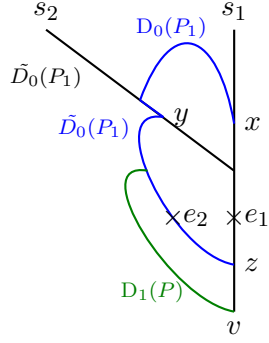

Figure 5: Shortest path avoiding $\{e_1, e_2\}$ is $P$. $D_1(P)$ last intersects $\tilde{P}_0(P_1) = P_0(s_2, v)$. $P_1$ diverges from $\tilde{P}_0(P_1)$ at $y$, i.e., $\tilde{D}_0(P_1) = P_1[y, z]$ (shown in blue). $P$ also diverges from $\tilde{P}_0(P)$ at $y$, i.e., $\tilde{D}_0(P) = P[y, v]$.

## 5 Extension to dual FT-MBFS

In this section we shall extend our analysis of the previous section to handle $\sigma$ sources using total $O(\sigma^{1/3} n^{5/3})$ space. We follow the approach similar to the case for single source. Let $S$ be the set of sources, where $|S| = \sigma$. Given a source $s$, let $\pi_s \subseteq \pi$ denote the ordering of edge failure of size upto 2. Let $\pi_s(0), \pi_s(1)$ and $\pi_s(2)$ be the subset of $\pi_s$ of size 0, 1 and 2 respectively. Our algorithm for finding dual FT-MBFS mimics the single source case.

---

**Procedure** Dual-FT-MBFS($S$,$v$,$\pi$): Augments the dual FT-MBFS subgraph $H$, such that for BFS tree of $G$ rooted at each $s \in S$ after any two edge failures in $G$, the incoming edges to $v$ are preserved in $H$.

**1 foreach** $s \in S$ **do** Dual-FT-BFS($s$,$v$,$\pi_s(0)$) ;
**2 foreach** $s \in S$ **do** Dual-FT-BFS($s$,$v$,$\pi_s(1)$) ;
**3 foreach** $s \in S$ **do** Dual-FT-BFS($s$,$v$,$\pi_s(2)$) ;

---

The first for loop in the above procedure finds shortest path from each source to $v$. We shall refer to the set of the shortest paths from each source to $v$ for different $s \in S$ as $\mathscr{P}_0$. We then move on to find the shortest path from each source to $v$ avoiding one edge failure and two failures respectively.

In the previous section, for each contributing path $P$ (that avoids $\geq 1$ failure), we saw that it necessarily diverges from $P_0$. Since we have multiple paths in $\mathscr{P}_0$, we define some new notations (see Figure 5).

**Definition 13.** (*Modified $P_0$ and $D_0(P)$*)

1. For any path $P$ (or its corresponding $P_1$), we define $\tilde{P}_0(P)$ (or $\tilde{P}_0(P_1)$) to be the last path from $\mathscr{P}_0$ which intersects with $P$ (or $P_1$), say at vertex $y$, such that at least one of $e_1$ or $e_2$ is present in $\tilde{P}_0(P)[y, v]$ (or $e_1 \in \tilde{P}_0(P_1)[y, v]$).

2. For any path $P$ (or $P_1$), we define $\tilde{D}_0(P) = P[v^*, v]$ (or $\tilde{D}_0(P_1) = P_1[v_0^*, v]$), where $v^*$ is the last vertex of $P$ (or $P_1$) on $P \cap \tilde{P}_0(P)$ (or $P_1 \cap \tilde{P}_0(P_1)$).

Note that in the single source case, both $P$ and $P_1$ diverge from the same path $P_0$. However, in multiple source that path $\tilde{P}_0(P)$ and $\tilde{P}_0(P_1)$ may differ. This is one of the major changes from the single source case. In fact, the reader will see that all our lemmas in Section 4 extend here with $P_0$ changed to $\tilde{P}_0(P)$ and $D_0(P)$ changed to $\tilde{D}_0(P)$. However, for completeness we have re-proven all lemmas.

We shall first extend the result of [15] (described in Lemma 28) for multiple sources in the following section.

## 5.1 Multiple failures on $\mathscr{P}_0$

For this we shall first present a property of $\tilde{D}_0(P)$ that shall be crucially used henceforth.

**Lemma 14.** *Let $P, P'$ be two contributing paths avoiding multiple failures only on paths in $\mathscr{P}_0$. Then $\tilde{D}_0(P) \cap \tilde{D}_0(P') = \{v\}$.*

*Proof.* We shall prove it by contradiction as follows. Let $w$ be the last vertex at which $\tilde{D}_0(P)$ intersects $\tilde{D}_0(P')$, after which they diverge to have two edge disjoint paths $\tilde{D}_0(P)[w,v]$ and $\tilde{D}_0(P')[w,v]$. Since both $P$ and $P'$ are contributing, $\tilde{D}_0(P)$ and $\tilde{D}_0(P')$ cannot intersect a path in $\mathscr{P}_0$ (else they will follow that path and be non contributing). Hence, being the preferred path, $\tilde{D}_0(P)[w,v]$ is the lexicographically shortest path avoiding $\mathscr{P}_0$. Similarly, $P'$ also being the preferred path $\tilde{D}_0(P')[w,v]$ also is the lexicographically shortest path avoiding $\mathscr{P}_0$. Thus, $P$ and $P'$ would merge after meeting at $w$, i.e., $\tilde{D}_0(P)[w,v] = \tilde{D}_0(P')[w,v]$, making one of them non-contributing. Hence our assumption is false proving the claim. □

This also establishes the following corollary.

**Corollary 15.** *Let $P, P'$ be two contributing paths avoiding multiple edge failures only on path in $\mathscr{P}_0$. Then paths $\tilde{D}_0(P)$ and $\tilde{D}_0(P')$ starts from different vertices on $\mathscr{P}_0$.*

Now, consider any source $s \in S$. Let $P_0 \in \mathscr{P}_0$ be the shortest path from $s$ to $v$. By Corollary 15 at most one contributing path $P$ can have $\tilde{D}_0(P)$ starting from each vertex on $P_0$, where each $\tilde{D}_0(P)$ would be disjoint. Let $P_0 = \{v = v_0, v_1, \cdots, v_k = s\}$. Since $\tilde{D}_0(P)$ starting from some $v_i \in P_0$ is a detour from the original path, $|\tilde{D}_0(P)| \geq |P_0[v_i, v]| = i$. Hence, using Lemma 14, $\tilde{D}_0(P)$ starting from $v_i$ can be uniquely associated with $i$ vertices. Now, for each $s \in S$, the paths with $\tilde{D}_0(P)$ initiating from some $v_i \in P_0$ with $i < \sqrt{n_\sigma}$ are limited to $\sqrt{n_\sigma}$ (using Corollary 15), having overall $\sqrt{n_\sigma} * \sigma = O(\sqrt{n\sigma})$ such paths. However, if $\tilde{D}_0(P)$ initiates from some $v_i \in P_0$ with $i \geq \sqrt{n_\sigma}$, it can be uniquely associated with $\sqrt{n_\sigma}$ of the $n$ vertices in the graph, limiting such paths to $n/\sqrt{n_\sigma} = O(\sqrt{n\sigma})$. Hence, for each $v \in V$ we have $O(\sqrt{n\sigma})$ contributing paths giving overall $O(\sqrt{\sigma}n^{3/2})$ paths. Thus, we get the following theorem.

**Theorem 16.** *The number of contributing paths where $F \in \mathscr{P}_0$ for any $s \in S$, where $|S| = \sigma$ are $O(\sqrt{\sigma}n^{3/2})$.*

## 5.2 Properties of Contributing paths

We now describe important properties of paths in $\mathscr{P}_0$ and contributing paths as follows.

**Lemma 17.** *The set of paths $\mathscr{P}_0$ is converging.*

*Proof.* Both $P_0$ and $P_0'$ are shortest paths from the corresponding source to $v$ without the requirement to avoid any failed edges. Hence, both would follow the shortest path from $w$ to $v$. Hence, they do not diverge after intersection giving $P_0[w,v] = P_0'[w,v]$. □

The number of contributing paths avoiding failures in $\mathscr{P}_0$ can easily bounded to $O(\sqrt{\sigma n})$ for each $v$ using Lemma 16. Excluding these paths, every contributing path satisfies the following properties.

**Lemma 18.** *Excluding $O(\sqrt{\sigma n})$ paths, for any contributing path $P$ from $s$ to $v$ avoiding $\{e_1, e_2\}$, the following properties holds true*

$\mathbb{P}_1$ : $e_1 \in P_0$ *and* $e_2 \in \mathrm{D}_0(P_1)$.

$\mathbb{P}_2$ : $\tilde{D}_0(P)$ *does not intersect with any path in $\mathscr{P}_0$. Also, if $\tilde{D}_0(P)$ diverges from $P_1$ it does not intersect it again.*

*Proof.* $\mathbb{P}_1$ : This property follows from Lemma 16.

$\mathbb{P}_2$ : By definition, $\tilde{D}_0(P)$ cannot pass through any path in $\mathscr{P}_0$. We will now prove the second part of $\mathbb{P}_2$. Consider a path $P$ with $\tilde{D}_0(P)$ intersecting $P_1$ after at some vertex $w$ after diverging from it at vertex $x \in P_1$. Again, $e_2 \in P_1[x,w]$ else $P$ would have taken $P_1[x,w]$ instead of $\tilde{D}_0(P)$. Hence, $e_2 \notin P_1[w,v]$ where we also have $e_1 \notin P_1$ (by definition of $P_1$). Thus, $P$ would continue to follow $P_1$ after $x$ making it non-contributing. □

## 5.3 Space Analysis

As described earlier, in order to bound the size of dual FT-MBFS subgraph to $O(\sigma^{1/3} n^{5/3})$, it suffices to bound the number of *contributing* paths from $s \in S$ to each vertex $v \in V$ avoiding two edge failures to $O(\sigma^{1/3} n^{2/3})$. Further, using $\mathbb{P}_1$ we are only concerned with a contributing path $P$ if $e_1 \in P_0$ and $e_2 \in \mathrm{D}_0(P_1)$. For the sake of highlighting similarity with single source case, we shall use $n_\sigma = n/\sigma$ throughout the section.

We first divide the paths in $\mathscr{P}_0$ into two parts as follows. For each $s \in S$, let $P_0(s,v)$ be the shortest path from $s$ to $v$. Let $v_{ls}$ be the vertex such that $|P_0(s,v)[v_{ls},v]| = n_\sigma^{1/3}$. We define $\mathcal{P}_{low} = \{P_0(s,v)[v_{ls},v] \mid s \in S\}$ and $\mathcal{P}_{high} = \{P_0(s,v)[s,v_{ls}] \mid s \in S\}$. This definition naturally extends the $P_{low}$ and $P_{high}$ defined in the single source case.

With this modified $\mathcal{P}_{low}$ and $\mathcal{P}_{high}$, we use the same definition of *standard paths* and hence $\mathscr{P}_a$ and $\mathscr{P}_{1a}$. However, the distinction of *long standard paths* ($\mathscr{P}_b$) from *short standard paths* ($\mathscr{P}_c$) would now be done by using $\tilde{D}_0(P_1)$ instead of $D_0(P_1)$. Hence, the *long standard paths* would be the standard paths with $|\tilde{D}_0(P_1)| \geq n_\sigma^{2/3}$. Finally, the definition of $LastPath(P)$ and $LastLeg(P)$ does not change, except in case $LastPath(P) = \phi$, we use $LastPath(P) = \tilde{P}_0(P)$ instead of $LastPath(P) = P_0$ (recall Definition 9). Moreover, the properties of $LastLeg(P)$ also remain same except for Lemma 10b which is modified as follows.

**Lemma 10.** *For every set $\mathscr{P}_x$ (for $x = a, b$ or $c$), we have the following.*

$b^*$. *Number of paths $P \in \mathscr{P}_x$ with $|LastLeg(P)| > n_\sigma^{\frac{1}{3}}$ or $LastPath(P) = \tilde{P}_0(P)$, is $O(\sigma^{\frac{1}{3}} n^{\frac{2}{3}})$.*

*Proof.* $b^*$. Consider any path $P \in \mathscr{P}_x$ with $|LastLeg(P)| > n_\sigma^{1/3}$. Using Lemma 10a, for each such $P$ we can associate $|LastLeg(P)| > n_\sigma^{1/3}$ unique vertices of $LastLeg(P)$ from a total of at most $n$ vertices. Thus, the number of such paths is $O(\sigma^{1/3} n^{2/3})$. Now, consider any path $P \in \mathscr{P}_x$ where $LastLeg(P)$ diverges from $\tilde{P}_0(P)$, i.e., $LastPath(P) = \tilde{P}_0(P)$. Using Lemma 10a, each such path emerges from a different vertex on $\mathscr{P}_0$, limiting the number of paths emerging from $\mathcal{P}_{low}$ to $O(\sigma * n_\sigma^{1/3}) = O(\sigma^{2/3} n^{1/3}) = O(\sigma^{1/3} n^{2/3})$. For the remaining paths, the corresponding $LastLeg(P)$ are at least as long as corresponding segment on $\mathcal{P}_{low}$, i.e., $n_\sigma^{1/3}$, limiting them to $O(\sigma^{1/3} n^{2/3})$ as described above. □

Equipped with these properties we can easily analyze the number of *non-standard paths* ($\mathscr{P}_a$) and *standard paths* ($\mathscr{P}_b$ and $\mathscr{P}_c$) in the following sections.

### 5.3.1 Analyzing non-standard paths $\mathscr{P}_a$ for dual FT-MBFS

Using Lemma 10b*, we know that the number of $P \in \mathscr{P}_a$ with $|LastLeg(P)| > n_\sigma^{1/3}$ or $LastPath(P) = \tilde{P}_0(P)$ is $O(\sigma^{1/3} n^{2/3})$. We now focus on the case when $|LastLeg(P)| \leq n_\sigma^{1/3}$ and $LastPath(P) \in \mathscr{P}_{1a}$. For any path $P$, let $v^* = Source(LastLeg(P))$. Since $LastLeg(P)$ is a detour from $LastPath(P)[v^*, v]$ avoiding the entire $P_0$ (using $\mathbb{P}_2$), we have $|LastPath(P)[v^*, v]| \leq |LastLeg(P)| \leq n_\sigma^{1/3}$. By definition, a contributing path $P$ is *non-standard* if either (a) $e_1 \in \mathcal{P}_{low}$, or (b) $D_0(P_1)$ merges with $P_0$ on $\mathcal{P}_{high}$, i.e., $Dest(D_0(P_1)) \in \mathcal{P}_{high}$. Hence, for every $P$, $LastPath(P)$ would correspond to one of the two cases (a) or (b). Case (b) is clearly not be applicable here because $|LastPath(P)[v^*, v]| \geq |\mathcal{P}_{low}| = n_\sigma^{1/3}$ (since $Dest(LastPath(P)) \in \mathcal{P}_{high}$). For case (a), on each $LastPath(P) \in \mathscr{P}_{1a}$, $v^*$ can be one of $n_\sigma^{1/3}$ vertices of $LastPath(P)$ closest to $v$. Further, since $e_1 \in \mathcal{P}_{low}$, there are only $n_\sigma^{1/3}$ such paths in $\mathscr{P}_{1a}$ because each such path corresponds to failure of unique edge in $\mathcal{P}_{low}$. Thus, since $|\mathcal{P}_{low}| = \sigma$, there are only $\sigma * n_\sigma^{1/3} \times n_\sigma^{1/3} = \sigma^{1/3} n^{2/3}$ different vertices $v^*$ limiting the number of $P \in \mathscr{P}_a$ with $|LastLeg(P)| \leq n_\sigma^{1/3}$ to $O(\sigma^{1/3} n^{2/3})$ (using Lemma 10a).

### Properties of standard paths ($\mathscr{P}_b$ and $\mathscr{P}_c$)

Recall the properties of *standard paths* described in Lemma 11. For multiple sources, Lemma 11a does not hold, because for two paths $P$ and $P'$, their corresponding paths $D_0(P_1)$ and $D_0(P_1')$ can diverge after intersection, if they start from different sources say $s_1, s_2$ (for $s_1 = s_2$, Lemma 11a applies). For example (see Figure 5), $P_1$ avoids $e_1$ on $P_0[s_1, v]$. Also, $D_0(P_1)$ passes through $P_0[s_2, v]$. Let $P_1'$ be a path avoiding $e_1'$ on $P_0[s_2, v] \cap P_1$, such that $D_0(P_1')$ intersect $D_0(P_1)$ before $D_0(P_1)$ enters $P_0[s_2, v]$. Hence, $D_0(P_1')$ has to diverge from $D_0(P_1)$ as $D_0(P_1)$ passes through $e_1'$ after the intersection.

This is the primary reason for defining *modified detour* $\tilde{D}_0(P_1)$, for which a lemma equivalent to Lemma 11a holds. Thus, the analysis of *standard paths* for multiple sources, uses $\tilde{D}_0(P_1)$ instead of $D_0(P_1)$ satisfying the following properties.

**Lemma 19.** *For the set of contributing* standard *paths, we have the following properties.*

a. *The set of paths* $\{\tilde{D}_0(P_1) | P_1 \in \mathscr{P}_{1b} \cup \mathscr{P}_{1c}\}$, *is converging.*

b. *The number of paths* $P \in \mathscr{P}_b \cup \mathscr{P}_c$, *which* $Source(LastLeg(P)) \notin \tilde{D}_0(P_1')$ *for some* $P_1' \in \mathscr{P}_{1b} \cup \mathscr{P}_{1c}$ *are* $O(\sigma^{1/3} n^{2/3})$.

*Proof.* a. For contradiction, assume that the last vertex at which $\tilde{D}_0(P_1)$ intersects $\tilde{D}_0(P_1')$ be $w$, after which they diverge. Since both $P_1$ and $P_1'$ are preferred shortest paths, it is only possible if $e_1 \in \tilde{D}_0(P_1')[w, v]$ (or $e_1' \in \tilde{D}_0(P_1)[w, v]$). Further, by definition of *standard paths*, both $e_1, e_1' \notin \mathcal{P}_{low}$ and $P_1, P_1'$ intersects $\mathscr{P}_0$ only on $\mathcal{P}_{low}$. Hence our assumption is false proving the claim.

b. Let $LastLeg(P)$ initiate from some vertex $w \in P_1'$, where $w \notin \tilde{D}_0(P_1')$. This implies that $P_1'[w, v]$ will intersect some $P_0(s'', v)$ at vertex $w'$ and diverges from it (since $w \notin \tilde{D}_0(P_1')$). Hence $e_1' \in P_0(s'', v)[w', v]$ and thus $w' \in \mathcal{P}_{high}$ (as $e_1' \in \mathcal{P}_{high}$ for *standard paths*). Further, since $P_1'[w', v]$ is a detour from $P_0(s'', v)[w', v]$ avoiding $P_0(s'', v)[v_{ls''}, v]$, we have $|P_1'[w', v]| \geq |P_0(s'', v)[v_{ls''}, v]| \geq n_\sigma^{1/3}$. Also, since $LastLeg(P)$ is a detour from $P_1'$ avoiding $|P_1'[w, v]|$, we have $|LastLeg(P)| \geq |P_1'[w, v]| \geq |P_1'[w', v]| \geq n_\sigma^{1/3}$. Using Lemma 10b*, we can thus limit the number of such paths to $O(\sigma^{1/3} n^{2/3})$. □

Using Lemma 19b, we only have to bound the number of *standard paths* whose $LastLeg(P)$ originates from some $\tilde{D}_0(P_1')$. Using Lemma 10b* and by definition of long standard paths $\mathscr{P}_b$, Theorem 12 is applicable for the set $\tilde{D}_0(P_1)$ for $P_1 \in \mathscr{P}_{1b}$ and $\alpha = n_\sigma^{1/3}$, bounding number of such paths in $\mathscr{P}_b$ to $O(\sigma^{1/3} n^{2/3})$. This leaves only the number of *short standard paths* that originate from some $\tilde{D}_0(P_1')$ described in the following section.

### 5.3.2 Analyzing short standard paths $\mathscr{P}_c$

Again, we only analyze the paths in $\mathscr{P}_c$ for undirected graphs here (extension for directed graphs is presented in Section 6). Using Lemma 10b* and Lemma 19b, we know that the number of $P \in \mathscr{P}_c$ with $|LastLeg(P)| > n_\sigma^{1/3}$ or $LastPath(P) = \tilde{P}_0(P)$ or $Source(LastLeg(P)) \notin \tilde{D}_0(P_1')$ (for some $P_1' \in \mathscr{P}_{1c}$) is $O(\sigma^{1/3} n^{2/3})$. We thus focus on the case when $|LastLeg(P)| \leq n_\sigma^{1/3}$ and $LastPath(P) \in \mathscr{P}_{1c}$ with $Source(LastLeg(P)) \in \tilde{D}_0(LastPath(P))$. Any such path can be divided into three parts (not necessarily non-empty) including (a) $P[s,x] = P_1[s,x]$, where $x = Source(\tilde{D}_0(P_1))$, (b) $P[x,y]$ where $y = Source(LastLeg(P))$ and (c) $P[y,v] = LastLeg(P)$.

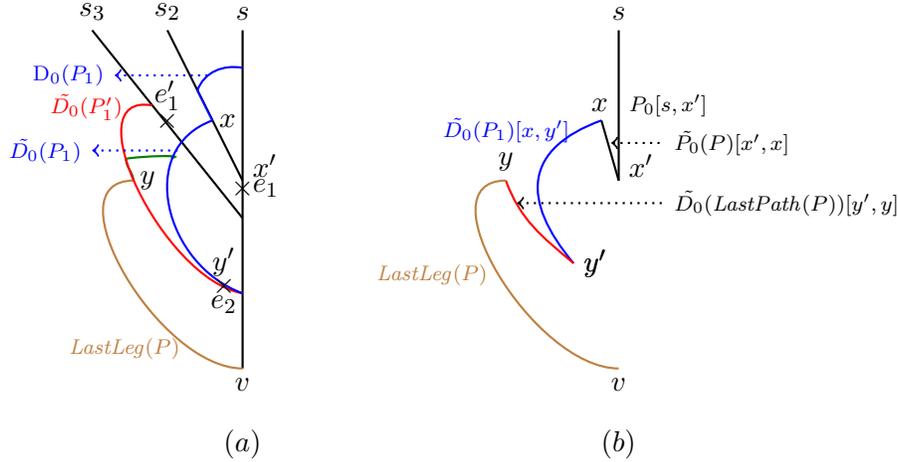

Figure 6: (a) Here $\tilde{P}_0(P_1) = P_0(s_2, v)$, $LastPath(P) = P_1'$ and $\tilde{P}_0(P_1') = P_0(s_3, v)$. (b) The path $P_0[s, x'] \cup \tilde{P}_0(P_1)[x', x] \cup \tilde{D}_0(P_1)[x, y'] \cup \tilde{D}_0(P_1')[y', y] \cup LastLeg(P)$ is a valid path avoiding $\{e_1, e_2\}$.

We find alternate paths for $P[s,x]$ and $P[x,y]$, which will help us in bounding their respective lengths (see Figure 6). By definition $\tilde{P}_0(P_1)$ (path from $s_2$ to $v$ in figure) intersects with $P_0$ and passes through $e_1$. Further, using Lemma 17 we know that $P_0$ and $\tilde{P}_0(P_1)$ will merge after the intersection at some point, say $x'$, where $e_1 \in \tilde{P}_0(P_1)[x', v] = P_0[x', v]$. Hence, we have an alternate path for $P_1[s, x]$ avoiding $e_1$ and $e_2$ (since $e_2 \notin \mathscr{P}_0$ by $\mathbb{P}_2$) formed by $P_0[s, x'] \cup \tilde{P}_0(P_1)[x', x]$. To bound $P[x, y]$, we first show that $\tilde{D}_0(P_1)$ and $\tilde{D}_0(LastLeg(P))$ intersect. Since $P$ diverges from $P_1$ on $\tilde{D}_0(P_1)$ (using Lemma 19b). and also diverges from $LastPath(P)$ (since $P$ is contributing), which reqirues $e_2 \in \tilde{D}_0(P_1)$ and $e_2 \in LastPath(P)[y, v]$ (as $e_1 \notin LastPath(P)$ because $LastPath(P)$ intersects on $\mathcal{P}_{low}$ and $e_1 \notin \mathcal{P}_{low}$ for *standard paths*). Moreover, since $P$ leaves $LastPath(P)$ before $e_2$, we have $e_2 \in \tilde{D}_0(LastPath(P))[y, v]$ (using Lemma 19b). Using Lemma 19a, we can thus say that $\tilde{D}_0(P_1)$ and $\tilde{D}_0(LastPath(P))$ merge at some vertex say $y'$, where $e_2 \in \tilde{D}_0(LastPath(P))[y', v] = \tilde{D}_0(P_1)[y', v]$ (see Figure 6(b)). We have an alternate path for $P[x, y]$ avoiding $F(P)$ formed by $\tilde{D}_0(P_1)[x, y'] \cup \tilde{D}_0(LastPath(P))[y', y]$. Since $P[s, v]$ is the shortest path avoiding $F(P)$ we have

$$
\begin{aligned}
|P| = |P[s,x]| + |P[x,y]| + |P[y,v]| \quad &= |P_0[s,x]| + |P_1[x,y]| + |P[y,v]| \\
&\leq (|P_0[s,x']| + |\tilde{P}_0(P_1)[x',x]|) + (|\tilde{D}_0(P_1)[x,y']| + |\tilde{D}_0(LastPath(P))[y',y]|) + |LastLeg(P)[y,v]| \\
&\leq |P_0[s,v]| + |\tilde{P}_0(P_1)[x,v]| + |\tilde{D}_0(P_1)[x,v]| + |\tilde{D}_0(LastPath(P))[y,v]| + |LastLeg(P)[y,v]| \\
&\leq |P_0[s,v]| + |\tilde{D}_0(P_1)[x,v]| + |\tilde{D}_0(P_1)[x,v]| + |\tilde{D}_0(LastPath(P))[y,v]| + |LastLeg(P)[y,v]| \\
&\qquad\qquad\qquad\qquad\qquad\qquad\qquad (\tilde{D}_0(P_1) \text{ is a detour from } \tilde{P}_0(P_1)) \\
&\leq |P_0[s,v]| + n_\sigma^{2/3} + n_\sigma^{2/3} + n_\sigma^{2/3} + n_\sigma^{1/3} \qquad\qquad \text{(by definition of } \mathscr{P}_c\text{)}
\end{aligned}
$$

Now, for any $s \in S$, let $\mathscr{P}_c(s)$ be the set of all contributing paths in $\mathscr{P}_c$ that start from $s$. Using Lemma 11b (that holds for $P \in \mathscr{P}_c(s)$), we know that for any $P, P' \in \mathscr{P}_c(s)$ we have $|P| \neq |P'|$. We thus arrange the paths in $\mathscr{P}_c(s)$ (except the ones in Lemma 10b* and Lemma 19b) in the increasing order of sizes, where $i^{th}$ such path has the length $\geq |P_0(s,v)| + i$ (as all paths at least as long as $P_0(s,v)$). Since for any such $P \in \mathscr{P}_c(s)$ we also have $|P| \leq |P_0[s,v]| + 4n^{2/3}$ (described above), clearly the number of paths in $\mathscr{P}_c(s)$ are $O(n_\sigma^{2/3})$ (for $i$ upto $4n^{2/3}$). Hence, overall the number of paths in $\mathscr{P}_c$ considering all sources $s \in S$ are $O(\sigma * n_\sigma^{2/3}) = O(\sigma^{1/3} n^{2/3})$.

# 6 Extending analysis of $\mathscr{P}_c$ for directed graphs

Our analysis of dual FT-BFS for undirected graphs extends to the directed case, except the case of short standard paths $\mathscr{P}_c$. The exact problem occurs when we use $LL(P)[y',y]$ when finding an alternate path in Section 4.2.3. While using this path, we crucially use the fact that the graph is undirected. Unfortunately, if the graph is directed, we cannot use this path. So, in this section, we give a different analysis for $\mathscr{P}_c$. We shall first define the notion of a *segmentable* paths (similar to the notion of *regions* used by Parter [15]) and describe its properties as follows.

**Definition 20.** *We define the following terms for any set of shortest paths $\mathcal{P}$ and $\mathcal{P}_*$ to the vertex $v$, where any $P \in \mathcal{P}$ is the shortest path from $Source(\mathcal{P}_*)$ avoiding failure of some edge $e \in P_*$ for some $P_* \in \mathcal{P}_*$. Also, let $\mathrm{D}_*(P)$ be the detour of $P \in \mathcal{P}$ from corresponding $P_* \in \mathcal{P}_*$.*

- **Segmentable Paths:** *The set of paths $\mathcal{P}_*$ is called segmentable if $\mathcal{P}_*$ is converging.*

- **Segments of $\mathcal{P}_*$:** *If any path $P_* \in \mathcal{P}_*$ intersects with other paths in $\mathcal{P}_*$ on vertices $x$ and $y$, the path $P_*[x,y]$ is called a segment of $\mathcal{P}_*$ if it does not intersect any path of $\mathcal{P}_*$ except at its endpoints. The segment of $\mathcal{P}_*$ from which $\mathrm{D}_*(P)$ of some $P \in \mathcal{P}$ emerges from the corresponding $P_* \in \mathcal{P}_*$ is represented by $P_s$.*

- **Representatives of a segment:** *For any segment $P_s = P_*[x,y]$, the shortest path from $x$ to $v$ avoiding $P_*[y,v]$ is called the* representative path *of $P_s$, and is denoted by $R_s$. Also, the last vertex of $R_s$ on $P_s \cap R_s$, say $v_{s^*}$, is called the* representative vertex *of $R_s$.*

The properties of a *segmentable* paths can be described as follows.

**Lemma 21.** *The segments of a set of segmentable paths $\mathcal{P}_*$ have the following properties.*

1. *Number of segments of $\mathcal{P}_*$ are $2|\mathcal{P}_*|$.*

2. *For any path $P' \in \mathcal{P}$ following the representative path of some segment $P_s$, $P'[v_{s^*}, v] = R_s[v_{s^*}, v]$.*

3. For any paths $P, P' \in \mathcal{P}$ that diverge from segment $P_s = P_*[x, y]$ and $P'_s = P'_*[x', y']$ respectively, if $P$ and $P'$ do not follow any path in $\mathcal{P}_*$ or representative path of any segment in $\mathcal{P}_*$, say set of such paths is $\mathcal{P}'$, we have

   (a) $e \in P_s[x, v_{s*}]$.
   
   (b) $\mathrm{D}_*(P)$ does not intersect with $P'_s[x', v'_{s*}]$ if $P'_s \neq P_s$.
   
   (c) The set of paths $\{\mathrm{D}_*(P) | P \in \mathcal{P}'\}$ is converging.

*Proof.* 1. By definition of *segmentable* paths, whenever two or more paths $P_*$ and $P'_*$ intersect they merge to form a single path. Thus, each $P_*$ originates from a unique vertex. Also, since on intersection of two or more such paths they merge to form a single path, the number of paths reduces by at least one after the intersection. This limits the total number of intersection points among the paths in $\mathcal{P}_*$ to $|\mathcal{P}_*|$. Now, each segment can be uniquely associated with starting vertex which can be a source of corresponding path in $\mathcal{P}_*$ or an intersection point described above. Hence, we have only $2|\mathcal{P}_*|$ segments.

2. Now, for the segment $P_s = P_*[x, y]$, the representative path is the shortest path avoiding $P_*[y, v]$. Hence, any path $P'$ which follows the representative path leaving $P_s$ necessarily has $e' \in P_*[v_{s*}, v]$. Since $R_s$ avoids entire $P_*[v_{s*}, v]$ (containing $e$) and $P'$ is not required to avoid any other edge in $\mathcal{P} \cup \mathcal{P}_*$, $P'$ will not diverge from $R_s$ after $v_{s*}$. Hence, we have $P'[v_{s*}, v] = R_s[v_{s*}, v]$.

3. Finally, consider the case when a path $P$ does not follow the representative path of any segment of $\mathcal{P}_*$.

   (a) Since $P$ leaves the segment $P_s$, unless $e \in R_s$ it would follow the shortest path avoiding $P_*[y, v]$ (i.e. $R_s$). Also, by definition $e \in P_*$, hence $e \in P_* \cap R_s = P_s[x, v_{s*}]$.
   
   (b) We shall prove this by contradiction. Hence, let $P'_s$ be the last segment to which $\mathrm{D}_*(P)$ intersects in the corresponding $P'_s[x', v'_{s*}]$. Clearly, $P$ diverges from $P'_s$ else it would intersect the next segment $P''_s$ at $x''$ (where $y' = x''$) if $P''_s$ exists, otherwise if $P'_s$ is the last segment $P$ will follow the corresponding path $P'_* \in \mathcal{P}_*$. Both these cases contradicts our assumption, ensuring $P$ diverges from $P'_s$. Since $e \in P_s$ and $P_s \neq P'_s$, we surely have $e \notin P'_s$. Since $P$ diverges from $P'_*$ and $e \notin P'_s$ (as $e \in P_s \neq P'_s$), we have $e \in P'_*[y', v]$. This ensures $e \notin P'_{s*}$ as it avoids the entire path $P'_*[y', v]$. Since $P$ intersects $P'_*[x', v'_{s*}]$ and avoids $P'_*[y, v]$, it will follow the shortest path avoiding $P'_*[y', v]$, i.e. the representative path $P'_{s*}$ as $e \notin P'_{s*}$. This contradicts our assumption and hence proves our claim.
   
   (c) Again, we shall prove this by contradiction. Assume $w$ is the last vertex at which $\mathrm{D}_*(P)$ intersects $\mathrm{D}_*(P')$ after which they diverge. Without loss of generality assume $|P[w, v]| \leq |P'[w, v]|$. Now, since $P'$ diverges from $P$, i.e., avoids $P[w, v]$ which implies $e' \in P[w, v]$ (as $w \neq v$). Further, using 3(a) and 3(b), we know that $e \in P'_s[x', v'_{s*}]$ whereas $\mathrm{D}_*(P)$ cannot intersect with any $P'_s[x', v'_{s*}]$. Thus, we have $e' \notin P[w, v]$ disproving our assumption and proving our claim.
   
   $\square$

## 6.1 Analyzing short standard paths $\mathscr{P}_c$ for dual FT-BFS

Using Lemma 11a, we know that the set of $\{\mathrm{D}_0(P_1) | P_1 \in \mathscr{P}_{1c}\}$ forms a set of segmentable paths ($\mathcal{P}_*$). Since we consider only those paths in $\mathscr{P}_c$ which emerge from $\mathrm{D}_0(P_1)$ (using Lemma 10b), the set of paths $\mathrm{D}_0(P)$ for $P \in \mathscr{P}_c$ starts from the same vertex as the corresponding $\mathrm{D}_0(P_1)$. Also, by definition of *standard paths*, $D_0(P_1)$ does not intersect with $P_{high}$ and $e_1 \in P_{high}$. Thus, $D_0(P)$ is the shortest path from $Source(D_0(P_1))$ to $v$ avoiding $e_2 \in D_0(P_1)$. Hence, for all $P \in \mathscr{P}_c$, we can apply Lemma 21 using $\mathcal{P}_*$ as the set $\{D_0(P_1)|P_1 \in \mathscr{P}_{1c}\}$, $\mathcal{P}$ as the set $\{D_0(P)|P \in \mathscr{P}_c\}$, $D_*(P)$ as $D_1(P)$, and $e \in P_*$ as the corresponding $e_2 \in P_1$.

Now, by definition every path in $\mathscr{P}_{1c}$ we have $|D_0(P_1)| \leq n^{2/3}$. Thus, the total number of paths in $D_0(P_1)$ for all $P_1 \in \mathscr{P}_{1c}$ are $O(n^{2/3})$. This also limits the number of segments of $D_0(P_1)$ to $O(n^{2/3})$ (using Lemma 21 (1)). Since, two contributing paths enter $v$ using a unique edge, using Lemma 21 (2) there will be exactly one path in $\mathscr{P}_c$ that follows the representative path of each segment, and none that follow any path in $\mathscr{P}_{1c}$ (else such paths are non-contributing). Thus the number of paths in $\mathscr{P}_c$ that follow the representative path or a path in $\mathscr{P}_{1c}$ is limited to $O(n^{2/3})$. Now, except for these $O(n^{2/3})$ paths, any two paths in $P, P' \in \mathscr{P}_c$ also satisfies that $D_1(P)$ and $D_1(P')$ do not diverge if they intersect (using Lemma 21 3(c)). Since, each path in $\mathscr{P}_c$ is contributing it hence enter $v$ from a unique edge, its $D_1(P)$ cannot intersect with detour of any other path in $\mathscr{P}_c$. The number of such paths with $|D_1(P)| > n^{1/3}$, can clearly by restricted to $O(n^{2/3})$. Hence, for the remaining paths in $\mathscr{P}_c$ we have $|D_1(P)| < n^{1/3}$ giving the following corollary.

**Lemma 22.** *Except for $O(n^{2/3})$ paths in $\mathscr{P}_c$, for any paths $P \in \mathscr{P}_c$ we have $|D_1(P)| \leq n^{1/3}$.*

We shall now bound the number of paths in $\mathscr{P}_c$ to $O(n^{2/3})$. Let $\mathscr{P}'_c$ be the set of paths in $\mathscr{P}_c$ removing $O(n^{2/3})$ paths described in Lemma 23. Also, let $\mathscr{P}'_{1c}$ be the corresponding $P_1$ for each $P \in \mathscr{P}'_c$. Thus, for any path $P \in \mathscr{P}'_c$ we have $|D_1(P)| \leq n^{1/3}$ and $|D_0(P_1)| < n^{2/3}$ (by definition of $\mathscr{P}_c$). Further, for any path $P \in \mathscr{P}_c$ we have $|P| \geq |P_0|$ (since $P$ is a detour from $P_0$). Let $x = Source(D_0(P_1))$ and $y = Source(D_1(P))$. Thus, any path $P \in \mathscr{P}'_c$ can be divided into three parts as follows.

$$|P| = |P[s,x]| + |P[x,y]| + |P[y,v]|$$
$$= |P_0[s,x]| + |D_0(P_1)[x,y]| + |D_1(P)[y,v]| \quad \text{(by definition of } x \text{ and } y\text{)}$$
$$\leq |P_0| + |D_0(P_1)| + |D_1(P)| \leq |P_0[s,v]| + n^{2/3} + n^{1/3}$$

Using Lemma 11b, we know that for any $P, P' \in \mathscr{P}_c$ we have $|P| \neq |P'|$. We thus arrange the paths in $\mathscr{P}'_c$ in the increasing order of sizes, where $i^{th}$ such path has the length $\geq |P_0| + i$. Since for any $P \in \mathscr{P}'_c$ we also have $|P| \leq |P_0| + 2n^{2/3}$, clearly the number of paths in $\mathscr{P}'_c$ (and hence $\mathscr{P}_c$) are $O(n^{2/3})$.

## 6.2 Analyzing short standard paths $\mathscr{P}_c$ for dual FT-MBFS

Like the single source case, we want to bound the number of contributing paths in $\mathscr{P}_c$ using segmentable paths. We first take case of all the paths $P$ with $D_0(P_1) \geq n^{2/3}_\sigma$ as follows.

Using Lemma 17, we know that $\mathscr{P}_0$ and hence $\mathcal{P}_{high}$ is a set of segmentable paths. Moreover, each $P_1 \in \mathscr{P}_{1c}$ is the shortest path to $v$ avoiding $e_1 \in \mathcal{P}_{high}$. Thus, for any $P_1 \in \mathscr{P}_{1c}$, $D_0(P_1)$ emerges from $\mathcal{P}_{high}$ and merges back to some $P_0 \in \mathcal{P}_{low}$. Hence, we can use Lemma 21 using $\mathcal{P}_*$ as $\mathcal{P}_{high}$, $\mathcal{P}$ as $\mathscr{P}_{1c}$, $D_*(P)$ as $D_0(P_1)$, and $e \in P_*$ as the corresponding $e_1 \in P_0$. Note that by definition, the representative path of $P_s = P_*[x,y]$, that is $R_s$, avoids $P_*[y,v]$ where $P_* \in \mathcal{P}_{high}$. Note that $P_*$ is actually a path from some source $s$ to $v$ till $v_{ls}$. Thus, $P_*[y,v]$ exists and our definition of representative path is still valid.

Consider any segment $P_s \in P_0$. Using Lemma 21 (2), we claim that once a path in $P'_1 \in \mathscr{P}_{1c}$ intersect $P_s$ and follow representative path of $P_s$ (i.e., $R_s$), then it cannot diverge from $R_s$. Using Lemma 19(b), for each such $P'_1$, we are concerned with those $P' \in \mathscr{P}_c$, that starts from $\tilde{D}_0(P'_1)$. We now claim that $\tilde{D}_0(P'_1)$ lies on $R_s$. This is due to the fact that both these paths are same from the segment $P_s$. We can easily bound the number of paths in $\mathscr{P}_c$, that starts from $R_s$. Let $\mathscr{P}_e$ denote the set of these paths. Since $\mathscr{P}_e \subset \mathscr{P}_c$, Lemma 18 applies on all the paths in $\mathscr{P}_e$. We will now bound $\mathscr{P}_e$ in a way similar to $\mathscr{P}_a$. Using Lemma 10($c^*$), all the paths in $P \in \mathscr{P}_e$ with $LastLeg(P)) \geq n_\sigma^{1/3}$ can be bounded to $O(\sigma^{1/3} n^{2/3})$. For the rest of the paths, their last path have length $\leq n_\sigma^{1/3}$, thus they necessarily starts on the lowest $n_\sigma^{1/3}$ vertices of representative paths. Using Lemma 21 (1), since $|\mathscr{P}_0| = \sigma$, there are at most $2\sigma$ representative paths. Thus the total number of path $P \in \mathscr{P}_e$ with small last paths is $O(\sigma n_\sigma^{1/3}) = O(\sigma^{1/3} n^{2/3})$.

Consider a path $P_1' \in \mathscr{P}_{1c}$ that follows $P^* \in \mathcal{P}_{high}$ but not any representative path. However, such path are non-standard paths (as $P_1'$ merges at $\mathcal{P}_{high}$). So, paths which follow $P^* \in \mathscr{P}_{1c}$ do not exist in $\mathscr{P}_{1c}$.

For the remaining paths $P_1, P_1' \in \mathscr{P}_{1c}$ (that donot follow any representative paths and donot follow $P^* \in \mathcal{P}_{high}$), using Lemma 21 (3c), we claim that $D_0(P_1)$ and $D_0(P_1')$ does not diverge after intersecting. Recall that in general this property is only satisfied by $\tilde{D}_0(P_1)$ (using Lemma 19a) hence $\mathscr{P}_b$ only handles the paths with $|\tilde{D}_0(P_1)| > (n_\sigma)^{2/3}$. We can thus similarly bound the number of paths with $|D_0(P_1)| > (n_\sigma)^{2/3}$ to $O(\sigma^{1/3} n^{2/3})$. Thus, we have the following lemma.

**Lemma 23.** *Except for $O(\sigma^{1/3} n^{2/3})$ paths in $\mathscr{P}_c$, for any contributing path $P \in \mathscr{P}_c$, $P_1$ does not follow any representative path or any path in $\mathcal{P}_{high}$ and $D_0(P_1) \leq (n_\sigma)^{2/3}$*

We are now ready to apply Lemma 21, this time with different parameters. Using Lemma 19a, we know that for paths $P_1 \in \mathscr{P}_{1c}$, the set of $\tilde{D}_0(P_1)$ forms a set of segmentable paths ($\mathcal{P}_*$). Since we consider only those paths in $\mathscr{P}_c$ which emerge from $\tilde{D}_0(P_1)$ (using Lemma 19b), the set of paths $\tilde{D}_0(P)$ for $P \in \mathscr{P}_c$ starts from the same vertex as the corresponding $\tilde{D}_0(P_1)$. Also, by definition of $\mathscr{P}_c$, $D_0(P_1)$ does not intersect with $\mathcal{P}_{high}$ and $e_1 \in \mathcal{P}_{high}$. Thus, $\tilde{D}_0(P)$ is the shortest path from $Source(\tilde{D}_0(P_1))$ to $v$ avoiding $e_2 \in \tilde{D}_0(P_1)$. Hence, for all $P \in \mathscr{P}_c$, we can apply Lemma 21 using $\mathcal{P}_*$ as the set $\{\tilde{D}_0(P_1)| P_1 \in \mathscr{P}_{1c}\}$, $\mathcal{P}$ as the set $\{\tilde{D}_0(P) \mid P \in \mathscr{P}_c\}$, $D_*(P)$ as $D_1(P)$, and $e \in P_*$ as the corresponding $e_2 \in P_1$.

Now, by definition every path in $\mathscr{P}_{1c}$ we have $|\tilde{D}_0(P_1)| \leq n_\sigma^{2/3}$. Thus, considering all sources the total number of paths in $\tilde{D}_0(P_1)$ for all $P_1 \in \mathscr{P}_{1c}$ are $O(\sigma n_\sigma^{2/3})$. This also limits the number of segments of $\tilde{D}_0(P_1)$ to $O(\sigma^{1/3} n^{2/3})$ (using Lemma 21 (1)). Since, two contributing paths enter $v$ using a unique edge, using Lemma 21 (2) there will be exactly one path in $\mathscr{P}_c$ that follows the representative path. Also, a contributing path in $\mathscr{P}_c$ does not follow any path in $\mathscr{P}_{1c}$ (else such paths are non-contributing). Thus, number of paths in $\mathscr{P}_c$ that follow the representative path or a path in $\mathscr{P}_{1c}$ is limited to $O(\sigma^{1/3} n^{2/3})$. Now, except for these $O(\sigma^{1/3} n^{2/3})$ paths, any two paths in $P, P' \in \mathscr{P}_c$ also satisfies that $D_1(P)$ and $D_1(P')$ do not diverge if they intersect (using Lemma 21 3(c)). Since, each path in $\mathscr{P}_c$ is contributing it enters $v$ from a unique edge, $D_1(P)$ cannot intersect with detour of any other path in $\mathscr{P}_c$. Since remaining $D_1(P)$ are disjoint, the number of such paths with $|D_1(P)| > (n_\sigma)^{1/3}$, can by restricted to $O(\sigma^{1/3} n^{2/3})$. Hence, for the remaining paths in $\mathscr{P}_c$ we have $|D_1(P)| < n_\sigma^{1/3}$ giving the following corollary.

**Lemma 24.** *Except for $O(\sigma^{1/3} n^{2/3})$ paths in $\mathscr{P}_c$, for any paths $P \in \mathscr{P}_c$ we have $|D_1(P)| \leq n_\sigma^{1/3}$.*

We shall now bound the number of paths in $\mathscr{P}_c$ to $O(\sigma^{1/3} n^{2/3})$. Let $\mathscr{P}_c'$ be the set of paths in $\mathscr{P}_c$ removing $O(\sigma^{1/3} n^{2/3})$ paths described in Lemma 23 and Lemma 24. Let $\mathscr{P}_c'(s)$ be the subset of paths in $\mathscr{P}_c'$ that start from a source $s \in S$. Also, let $\mathscr{P}_{1c}'(s)$ be the corresponding $P_1$ for each $P \in \mathscr{P}_c'(s)$. Using Lemma 23 and Lemma 24, for any path $P \in \mathscr{P}_c'(s)$, we have $|D_0(P_1)| \leq n_\sigma^{1/3}$ and $|D_1(P)| \leq n_\sigma^{1/3}$.

For any path $P \in \mathscr{P}_c'(s)$, we have $|P| \geq P_0[s, v]$. Let $x = Source(D_0(P_1))$ and $y = Source(D_1(P))$.

Thus, $P$ can be divided into three parts as follows.

$$\begin{aligned}
|P| &= |P[s,x]| + |P[x,y]| + |P[y,v]| \\
&= |P_0[s,x]| + |\mathrm{D}_0(P_1)[x,y]| + |\mathrm{D}_1(P)[y,v]| \quad \text{(by definition of } x \text{ and } y\text{)} \\
&\leq |P_0[s,v]| + |\mathrm{D}_0(P_1)| + |\mathrm{D}_1(P)| \leq |P_0[s,v]| + n_\sigma^{2/3} + n_\sigma^{1/3}
\end{aligned}$$

Using Lemma 11b, we know that for any $P, P' \in \mathscr{P}_c(s)$ we have $|P| \neq |P'|$. We thus arrange the paths in $\mathscr{P}'_c(s)$ in the increasing order of sizes, where $i^{th}$ such path has the length $\geq |P_0[s,v]| + i$. Hence, the maximum value of such $i$ is $O(n_\sigma^{2/3})$ limiting total number of such paths in $\mathscr{P}'_c(s)$ to $O(n_\sigma^{2/3})$. Thus the total number of paths in $\cup_s \mathscr{P}'_c(s) = O(\sigma n_\sigma^{2/3}) = O(\sigma^{1/3} n^{2/3})$.

## 7 Dual Fault Tolerant Spanner with additive stretch 2

Fault tolerant spanners have been extensively studied considering only multiplicative stretch [8, 10], whereas for additive stretch it was first studied by Braunschvig et al.[7] supporting only edge faults. Parter [14] presented the first additive fault tolerant spanners under a single vertex failure, which was later improved by Bilo et al. [4]. Recently, Bodwin et al. [6] presented the first additive fault tolerant spanner to handle multiple failures requiring $O(fn^{2-\frac{1}{2f+1}})$ space for an $f$ fault tolerant spanner with additive stretch 2.

Bilo et al. [4] improved upon the construction by Parter [14] to use a fault tolerant sourcewise $\beta$ additive spanner to develop a fault tolerant $\beta + 2$ additive spanner as follows.

**Theorem 25** ([4]). *A $k$ fault $\sigma$ sourcewise $\beta$ additive spanner requiring $O(f(n,\sigma,k))$ edges, can be used to develop a $k$ fault $\beta + 2$ additive spanners requiring $O(f(n,\sigma,k) + n^2/\sigma)$ edges for any $1 \leq \sigma \leq n$.*

Since dual fault FT-MBFS from $\sigma$ sources is essentially a 2 fault $\sigma$ sourcewise $\beta = 0$ additive spanner, we have $f(n,\sigma,2) = n^{5/3}\sigma^{1/3}$. Hence, choosing $\sigma = n^{1/4}$ we get an algorithm to build a dual fault $+2$ additive spanner of size $O(n^{7/4})$. Our result thus improves the bounds for dual fault tolerant $+2$ spanner from $O(n^{9/5})$ [6] to $O(n^{7/4})$.

## 8 Conclusion

In this paper, we simplified the analysis in [15] for dual FT-BFS problem and extended it to dual FT-MBFS problem. Unfortunately, extending our result to $k$ FT-MBFS (or even $k$ FT-BFS) problem requires a lot of case analysis. Ideally, one would wish to design a simple data structure to handle multiple failures using some new insight with little or no case analysis. A natural step would be to completely understand these simple cases and derive significant inferences from them to develop new techniques. The simplicity of FT-BFS structure [16] enables a clear understanding of the basic technique used for its construction and analysis. Our work aims to be a significant step to achieve the same for dual FT-BFS by simplifying the result of [15] and generalizing it similar to [16].

## A  Multiple Failures on $P_0$

We shall now bound the number of contributing paths avoiding multiple failures on $P_0$ to $O(n^{3/2})$. The result was also proved by [15], we present it here for the sake of completeness and to highlight its simplicity using our approach. For this we shall first present a property of contributing paths which establishes that the *last leg* of two contributing paths cannot intersect except at $v$.

**Lemma 26.** *For any two contributing paths $P, P'$ avoiding multiple failures only on $P_0$, let $x, x'$ ($x, x' \neq v$) be the last vertex of $P$ and $P'$ respectively on $P_0$, then $P[x, v] \cap P'[x', v] = \{v\}$.*

*Proof.* We shall prove it by contradiction as follows. Let $w$ be the last vertex (except $v$) at which $P[x, v]$ intersects with $P'[x', v]$. Hence, $P[w, v]$ and $P'[w, v]$ are vertex disjoint except at $w$ and $v$. By definition of $x$, $P[w, v]$ does not intersects $P_0$. Thus, being the preferred path, $P[w, v]$ is the lexicographically shortest path avoiding $P_0$. Similarly, $P'$ also being the preferred path $P'[w, v]$ also is the lexicographically shortest path avoiding $P_0$. Further, $P$ and $P'$ would merge after meeting at $w$, i.e., $P[w, v] = P'[w, v]$, making one of them non-contributing. Hence, our assumption is false proving the claim. □

This also establishes the following corollary.

**Corollary 27.** *For any two contributing paths $P, P'$ avoiding multiple edge failures only on $P_0$, let $x, x'$ ($x, x' \neq v$) be the last vertex of $P$ and $P'$ respectively on $P_0$, then $x \neq x'$.*

Using Corollary 27 at most one contributing path $P$ can have $P[x, v]$ starting from each vertex on $x \in P_0$. Let $P_0 = \{s = u_k, u_{k-1}, \cdots, u_1 = v\}$. Thus, the number of paths with $x = u_i$ and $i \leq \sqrt{n}$ are limited to $O(\sqrt{n})$, each starting from different $u_i \in P_0$ with $i \leq \sqrt{n}$. Now, for each path $P$ where $P[x, v]$ starts from some $x = u_i \in P_0$, we have $|P[x, v]| \geq |P_0[u_i, v]| = i$ (as $P[x, v]$ is a detour from $P_0[x, v]$). Using Lemma 26, $P$ can be uniquely associated with at least $i$ vertices of $P[x, v]$. Thus, if $i \geq \sqrt{n}$, the number of such contributing paths are limited to $O(n/\sqrt{n}) = O(\sqrt{n})$ as each is associated to at least $\sqrt{n}$ different vertices. Hence, for each $v \in V$ we have $O(\sqrt{n})$ contributing paths giving overall $O(n^{3/2})$ paths. Thus, we get the following theorem.

**Theorem 28.** *The number of contributing paths $P$ with $\mathrm{F}(P) \in P_0$ are $O(n^{3/2})$.*

## B  Omitted Proofs

**Lemma 11.** *For the set of contributing* standard *paths, we have the following properties.*

a. *The set of paths $\{\mathrm{D}_0(P_1) | P_1 \in \mathscr{P}_{1b} \cup \mathscr{P}_{1c}\}$, is converging.*

b. *(Parter [15]) For any two paths $P, P' \in \mathscr{P}_b \cup \mathscr{P}_c$, $|P| \neq |P'|$.*

*Proof*.b. We shall proof this by contradiction, hence assume $|P| = |P'|$. Without loss of generality, assume that $e_1$ is at least as high as $e'_1$ on $P_0$. Thus, when $D_0(P)$ leaves $P_0$ to avoid $e_1$ it cannot intersect $P_0$ (and hence $e'_1$) using $\mathbb{P}_2$. Now, we have two cases, (a) $P$ does not pass through $e'_2$ as well, or (b) $P$ passes through $e'_2$. In case (a) $P$ is also a valid path for failure of $\{e'_1, e'_2\}$. Since $|P| = |P'|$, $P'$ can be contributing only if $P$ is not preferred over $P'$ for $\{e'_1, e'_2\}$, i.e., $P'$ leaves $P_0$ earlier than $P$. Hence, $P'$ also avoids $e_1$. Since both $P$ and $P'$ cannot avoid $F(P)$ and $F(P')$, $P'$ must pass through $e_2$. Thus, in case (a) $P'$ avoids $e_1$ and passes through $e_2$, and in case (b) $P$ avoids $e'_1$ and passes through $e'_2$. Hence, without loss of generality for two contributing *standard* paths $P^*, P^\circ$ with $|P^*| = |P^\circ|$ such that $P^*$ avoids $e_1^\circ$ and passes through $e_2^\circ$ (case (a): $\{P^*, P^\circ\} = \{P', P\}$, case (b): $\{P^*, P^\circ\} = \{P, P'\}$). We will now prove that both case (a) and (b) cannot occur thus proving the lemma.

For contradiction, assume that $P^*$ avoids $e_1^\circ$ and passes through $e_2^\circ = (x, y)$ (say). Thus, $P^*$ necessarily intersects $P_1^\circ[y, v]$ (as $e_2^\circ \in P_1^\circ$ using $\mathbb{P}_1$). Since $P^*$ is contributing, it has to leave $P_1^\circ[y, v]$ implying that either of $e_1^*$ or $e_2^*$ would lie on $P_1^\circ[y, v]$. By definition of *standard paths*, we know $e_1^* \notin P_1^\circ[y, v]$ implying $e_2^* \in P_1^\circ[y, v]$. Hence, both $e_2^*$ and $e_2^\circ$ lie on $P_1^\circ$, with $e_2^\circ$ closer to $s$ than $e_2^*$ on $P_1^\circ$ (see Figure 7). Moreover, using $\mathbb{P}_2$ we know that $P^\circ$ cannot intersect $P_1^\circ$ after $e_2^\circ$ and hence $P^\circ$ avoids $e_2^* \in P_1^\circ[y, v]$. Thus, $P^\circ$ also avoids the failure of $\{e_1^\circ, e_2^*\}$, where $\{e_1^\circ, e_2^*\} \prec_\pi \{e_1^\circ, e_2^\circ\}$ as $e_2^*$ lower than $e_2^\circ$ on $P_1^\circ$. Hence, using $\mathbb{P}_4$, we must have another shortest path $P''$ that is preferred over $P^\circ$ for $\{e_1^\circ, e_2^*\}$. There are following cases:

(a) $e_1^\circ = e_1^*$

In this case, $P'' = P^*$. Since $P^*$ is leaving $P^\circ$ lower than $P^\circ$ it is less preferred than $P^\circ$, leading to a contradiction.

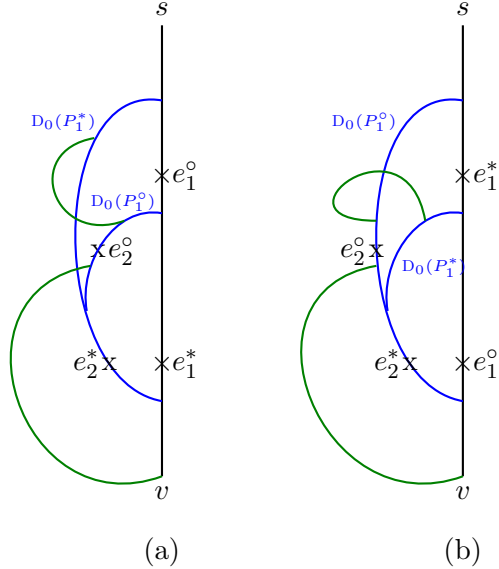

(a)          (b)

Figure 7: Pictorial representation of two cases referred to in the proof of Lemma 11b. The green path represents $P''$.

(b) $e_1^\circ \neq e_1^*$

Now, $P''$ cannot avoid $e_2^\circ$ otherwise $P''$ would be valid for failure of $F(P^\circ)$ leading to contradiction that $P^\circ$ is preferred path for $F(P^\circ)$ (since $P''$ preferred over $P^\circ$). Thus, $P''$ passes through $e_2^\circ \in P_1^\circ$ and is also preferred over $P^\circ$. Since $P^\circ$ avoids $e_2^* \in P_1^\circ$ this is possible only if $|P''| < |P^\circ| = |P^*|$. Further, after passing through $e_2^\circ$ both $P^*$ and $P''$ must follow the same shortest paths avoiding $e_2^* \in P_1^\circ$. Now, we consider two cases:

i. $\{e_1^*, e_2^*\} \prec_\pi \{e_1^\circ, e_2^*\}$ (see Figure 7(a))
   In this case, $P''$ leaves $P_0$ before $e_1^\circ$ and hence $e_1^*$, becoming valid for failure of F($P^*$). This makes $P^*$ non-contributing as $|P^*| < |P''|$.

ii. $\{e_1^\circ, e_2^*\} \prec_\pi \{e_1^*, e_2^*\}$ (see Figure 7(b))
   Since both $P^*$ and $P''$ follows the same path after passing through $e_2^\circ$, again $P^*$ becomes non-contributing as F($P''$) $\prec_\pi$ F($P^*$).

□